\documentclass[letter, traditabstract]{aa} 
\usepackage{graphicx}
\usepackage[varg]{txfonts}
\usepackage{upgreek} 
\usepackage{multirow} 
\usepackage[FIGTOPCAP, center, nooneline]{subfigure}
\usepackage[]{natbib}
\usepackage{rotating}
\usepackage{lscape}
\newcommand{\HII}{H\,{\sc ii}\,\,}

\bibpunct{(}{)}{;}{a}{}{,}

\usepackage{setspace} 

\usepackage{color} 
\begin{document}

   \title{Chemical segregation of complex organic O-bearing species \\ in Orion KL\thanks{This paper makes use of the following ALMA data:
ADS/JAO.ALMA\#2011.0.00009.SV. ALMA is a partnership of
ESO (representing its member states), NSF (USA), and NINS (Japan), together with NRC (Canada), NSC and ASIAA (Taiwan), and KASI (Republic of Korea), in cooperation with the Republic of Chile. The Joint ALMA Observatory is operated by ESO, AUI/NRAO, and NAOJ.}}

     \author{B. Tercero\inst{\ref{inst1}}\and S. Cuadrado\inst{\ref{inst2}}\and A. L\'opez\inst{\ref{inst2}} \and N. Brouillet\inst{\ref{inst3}}\and D. Despois\inst{\ref{inst3}} \and J. Cernicharo\inst{\ref{inst2}}}

   \institute{
   Observatorio Astron\'omico Nacional (OAN-IGN). Calle Alfonso XII, 3, E-28014 Madrid, Spain.\label{inst1}
   \and Instituto de F\'{\i}sica Fundamental (IFF-CSIC). Calle Serrano 123, E-28006 Madrid, Spain. \label{inst2}
   \and Laboratoire d'astrophysique de Bordeaux, Univ. Bordeaux, CNRS, B18N, allée Geoffroy Saint-Hilaire, 33615 Pessac, France. \label{inst3}\\
\email{b.tercero@oan.es; [s.cuadrado;jose.cernicharo]@csic.es}}

   \date{Received - 2018; accepted - 2018}

 
  \abstract{We investigate the chemical segregation of complex O-bearing species 
  (including the largest and most complex ones detected to date in space) 
  towards Orion KL, the closest high-mass star-forming region. The molecular line images obtained using the ALMA science verification data
  reveal a clear segregation of chemically related species depending
  on their different functional groups. We map the emission
  of $^{13}$CH$_3$OH, HCOOCH$_3$, CH$_3$OCH$_3$, CH$_2$OCH$_2$, CH$_3$COOCH$_3$, HCOOCH$_2$CH$_3$, CH$_3$CH$_2$OCH$_3$,
  HCOOH, OHCH$_2$CH$_2$OH, CH$_3$COOH, CH$_3$CH$_2$OH, CH$_3$OCH$_2$OH, OHCH$_2$CHO, and CH$_3$COCH$_3$ 
  with $\sim$1.5$''$  angular resolution and provide molecular abundances of these species toward different
  gas components of this region. We disentangle the emission of these species in the different Orion components
  by carefully selecting lines free of blending and opacity effects. Possible effects in the molecular spatial distribution
  due to residual blendings and different excitation conditions are also addressed.
  We find that while species containing the C$-$O$-$C group, i.e. an ether group,
  exhibit their peak emission and higher abundance towards the compact ridge,
  the hot core south is the component where species containing a hydroxyl group ($-$OH)
  bound  to a carbon atom (C$-$O$-$H) present their emission peak and higher abundance.
  This finding allows us to propose methoxy (CH$_3$O$-$) and hydroxymethyl ($-$CH$_2$OH) radicals
  as the major drivers of the chemistry in the compact ridge and the hot core south, respectively,
  as well as different evolutionary stages and
  prevailing physical processes in the different Orion components.
 }

   


   \maketitle
%
\section{Introduction}\label{Intro}

Orion~BN/KL (Becklin-Neugebauer\,/\,Kleinmann-Low) is the closest high-mass star-forming
region ($\sim$\,400\,pc; \citealt{Grossschedl_2018,Kounkel_2017,Menten_2007})
exhibiting several processes related to young stellar objects and high-mass star formation 
(see, e.g. \citealt{Genzel_1989}). This region is located at the core of the Orion Molecular Cloud~1 (OMC\,1),
which lies behind the Orion Nebula cluster \citep{ODell_2001}. 
The core of Orion~KL contains at least three self-luminous objects (protostars),
the compact \HII regions $I$, $n$, and $BN$; these sources are within a region of
$\sim$\,10$''$ ($\sim$\,0.02\,pc). Their proper motions reveal that they are moving away from
a common region \citep{Gomez_2005}. Different scenarios, as well as the formation 
of high-mass stars, have been proposed to explain the complexity of this source.
An explosion of a multi-star system (sources $I$, $n$, $x$, and $BN$) that took place $\sim$\,500
years ago \citep{Gomez_2005,Bally_2017,Luhman_2017,Rodriguez_2017} has been proposed as the main factor responsible for most of Orion~KL gas 
components \citep{Zapata_2011}.

The different gas components show distinct physical and chemical properties.
Classically, these components have been identified with single-dish telescopes by a 
characteristic systemic velocity \citep{Blake_1987,Schilke_2001,Tercero_2010}:
(i) hot core, a hot (\mbox{$T$\,$\simeq$\,200\,$-$\,300\,K}), dense clump rich in complex organic saturated N-bearing species such as
CH$_3$CH$_2$CN, characterised by 
\mbox{$\Delta v$\,$\simeq$\,5\,$-$\,15\,km\,s$^{-1}$} and \mbox{$v_{\rm LSR}$\,$\simeq$\,5\,$-$\,7\,km\,s$^{-1}$};
(ii) extended ridge, the host, quiescent, and relatively cold (\mbox{$T$\,$\simeq$\,60\,K}) ambient cloud rich in simple species such as CS or HCN,
emitting lines with \mbox{$\Delta v$\,$\simeq$\,3\,$-$\,4\,km\,s$^{-1}$} and \mbox{$v_{\rm LSR}$\,$\simeq$\,8\,$-$\,9\,km\,s$^{-1}$}; 
(iii) compact ridge, 
a warm (\mbox{$T$\,$\simeq$\,150\,K}), compact clump rich in organic saturated O-rich species such as HCOOCH$_3$ or CH$_3$OCH$_3$,
whose spectral features are characterised by 
\mbox{$\Delta v$\,$\simeq$\,2\,$-$\,3\,km\,s$^{-1}$} and \mbox{$v_{\rm LSR}$\,$\simeq$\,7\,$-$\,8\,km\,s$^{-1}$}; and 
(iv) plateau, molecular
outflows presenting typical shock chemistry with molecules such as SO or SiO \citep{Tercero_2011,Goicoechea_2015}; here the low-velocity flow 
is characterised by lines with \mbox{$\Delta v$\,$\simeq$\,20\,km\,s$^{-1}$} and \mbox{$v_{\rm LSR}$\,$\simeq$\,5\,$-$\,6\,km\,s$^{-1}$}
\citep{Bell_2014}, whereas the high-velocity flow
presents lines with \mbox{$\Delta v$ as wide as 150\,km\,s$^{-1}$} and \mbox{$v_{\rm LSR}$\,$\simeq$\,10\,km\,s$^{-1}$}.

In the recent years, data from the last generation of telescopes have added further complexity to this region.
\citet{Neill_2013} and \citet{Crockett_2014}
noted that a component between hot core and compact ridge,  called the hot core south
(\mbox{$\Delta v$\,$\simeq$\,5\,$-$\,10\,km\,s$^{-1}$} and \mbox{$v_{\rm LSR}$\,$\simeq$\,6.5\,$-$\,8\,km\,s$^{-1}$}) 
was required to correctly model the \textit{Herschel}-HIFI emission of several molecules. In these works, this component was also revealed
by mapping the emission of HDO, $^{13}$CH$_3$OH, $^{13}$CH$_3$CN, and HCOOCH$_3$ using the ALMA Science Verification (SV) data.
Moreover, \citet{Peng_2012} previously noted that the strongest CH$_2$DOH and CH$_3$OH emissions come from 
the hot core south-west region with a velocity that is typical of the compact ridge.

This new component is especially important regarding the spatial distribution of the complex organic O-bearing molecules.
A series of recent works based on interferometric data have demonstrated a clear spatial differentiation
between the most complex O-bearing species detected in Orion~KL.
On the one hand, the following species (among others) emit from the compact ridge 
component: 
HCOOCH$_3$ \citep{Favre_2011a,Widicus_2012,Crockett_2014,Tercero_2015,Cernicharo_2016}, 
CH$_3$OCH$_3$ \citep{Favre_2011b,Widicus_2012,Brouillet_2013,Feng_2015,Tercero_2015,Cernicharo_2016}, 
CH$_3$OH \citep{Friedel_2012,Peng_2012,Neill_2013,Feng_2015,Tercero_2015,Cernicharo_2016}, 
CH$_3$CH$_2$OCH$_3$ \citep{Tercero_2015}, 
HCOOCH$_2$CH$_3$ \citep{Tercero_2015}, and 
CH$_2$OCH$_2$ \citep{Cernicharo_2016}.
On the other hand, complex organic O-rich species, whose major emission is associated with
the hot core south, emit far away the compact ridge: 
CH$_3$COCH$_3$ \citep{Widicus_2012,Peng_2013,Feng_2015,Cernicharo_2016},
OHCH$_2$CH$_2$OH \citep{Brouillet_2015,Cernicharo_2016,Favre_2017},
CH$_3$COOH \citep{Cernicharo_2016,Favre_2017}, 
CH$_3$CH$_2$OH \citep{Feng_2015,Tercero_2015,Cernicharo_2016}, and 
OHCH$_2$CHO \citep{Cernicharo_2016}. 
\citet{Favre_2017} proposed that the rich molecular composition of the 
Orion core, which exhibits the emission peak of OHCH$_2$CH$_2$OH and CH$_3$COOH,
may reflect gas-phase chemistry in an induced shock or
post-shock stage due to the impact of a bullet of matter ejected during the explosive event
of Orion~KL \citep{Wright_2017}.

Chemical segregation of complex organic molecules (COMs, \citealt{Herbst_2009})
in star-forming regions has been investigated
concerning the spatial differentiation between the O-~and N-bearing
species \citep{Peng_2013,Oberg_2013,Fayolle_2015,JimenezSerra_2016,Allen_2017,Bergner_2017,Suzuki_2018}.
However, information about the spatial distribution of chemically related COMs
is lacking, mostly due to the low abundances of the most complex
species which prevent the detection of these species in the majority of
star-forming regions. The identification of several O-bearing COMs in Orion~KL
allows us to investigate the spatial distribution of similar species (in terms
of complexity and variety of atoms) but harbouring different chemical functional groups
to test the COM formation and evolution.

In this Letter we investigate the spatial distribution of O-bearing COMs and some related
species detected in this source based on the ALMA SV data,
including, for the first time, the largest and most complex O-bearing COMs detected  to date in space 
(Sect.\,\ref{results}): $^{13}$CH$_3$OH (methanol), HCOOCH$_3$ (methyl formate), 
CH$_3$OCH$_3$ (dimethyl ether), CH$_2$OCH$_2$ (oxirane), CH$_3$COOCH$_3$ (methyl acetate),
HCOOCH$_2$CH$_3$ (ethyl formate), CH$_3$CH$_2$OCH$_3$ (ethyl methyl ether), HCOOH (formic acid), OHCH$_2$CH$_2$OH
(ethylene glycol), CH$_3$COOH (acetic acid), CH$_3$CH$_2$OH (ethanol), CH$_3$OCH$_2$OH (methoxymethanol), OHCH$_2$CHO (glycolaldehyde), and CH$_3$COCH$_3$ (acetone).
In addition, molecular abundances have been derived for all these species in different regions of Orion.
Finally, we discuss the observed chemical differentiation in Sect.\,\ref{dis}. 



\section{Observations}\label{Obs}

The ALMA Science Verification (SV) data\footnote{http://almascience.eso.org/almadata/sciver/OrionKLBand6/} were taken in January~2012 towards the IRc2 region in Orion. The observations were carried out with 16~antennas of 12\,m in the frequency range from 213.715\,GHz to 246.627\,GHz (Band~6). The primary beam was $\sim$\,27$''$. Spectral resolution was 0.488\,MHz
($\sim$\,0.64\,km\,s$^{-1}$ in the observed frequency range). The observations were centred on \mbox{$\upalpha_{\rm J2000}$\,=\,05$^{\rm h}$\,35$^{\rm m}$\,14.35$^{\rm s}$}, \mbox{$\updelta_{\rm J2000}$\,=\,$-$05$^{\circ}$\,22$'$\,35.00$''$}.
The CASA software\footnote{http://casa.nrao.edu} was used for initial processing, and then the visibilities were exported to the GILDAS package\footnote{http://www.iram.fr/IRAMFR/GILDAS}
for further analysis. The line maps were cleaned using the
HOGBOM algorithm \citep{Hogbom_1974}. The synthesised beam
ranged from \mbox{2.00$''$\,$\times$\,1.48$''$} with a position angle~(PA) of 176$^{\circ}$ at 214.0\,GHz to
\mbox{1.75$''$\,$\times$\,1.29$''$} with a PA of 164$^{\circ}$ at 246.4\,GHz. The brightness
temperature to flux density conversion factor is 9\,K for 1\,Jy per beam. The continuum emission was subtracted in the maps by carefully selecting line-free channels.

\begin{figure*}
\vspace*{3cm}
\begin{minipage}[t]{0.2\textwidth}
\begin{center}
\vspace*{-3cm}\includegraphics[scale=0.55,angle=0]{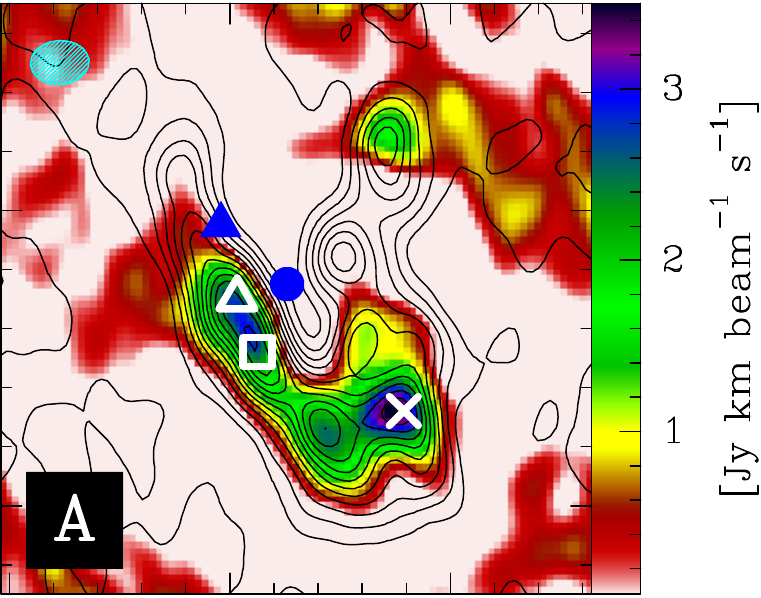}\vspace{0.2cm}
\includegraphics[scale=0.55,angle=0]{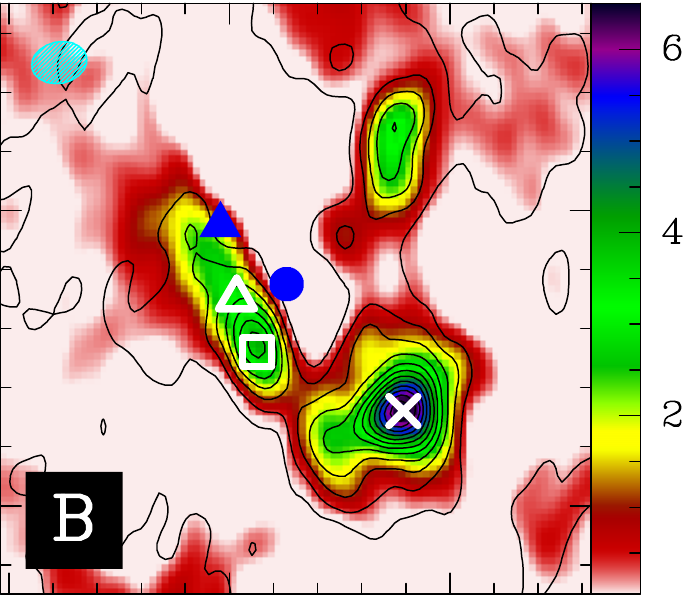}\vspace{0.13cm}\\
\includegraphics[scale=0.55,angle=0]{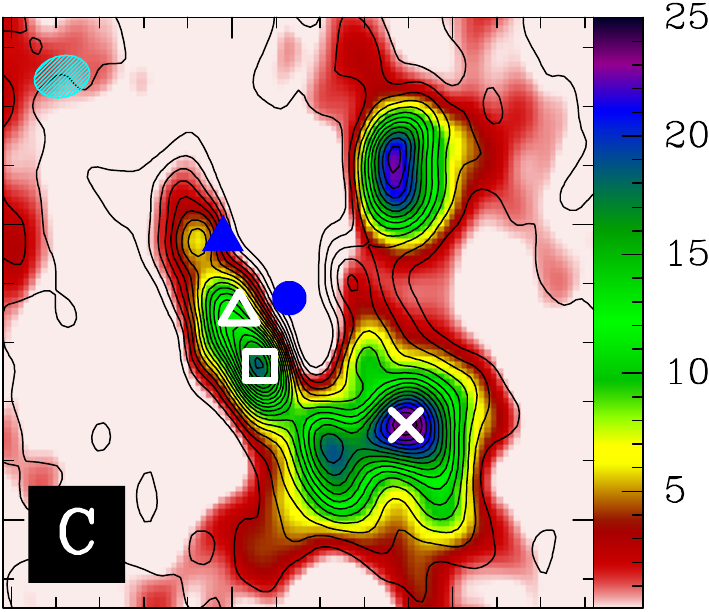}\vspace{0.2cm}\\
\includegraphics[scale=0.55,angle=0]{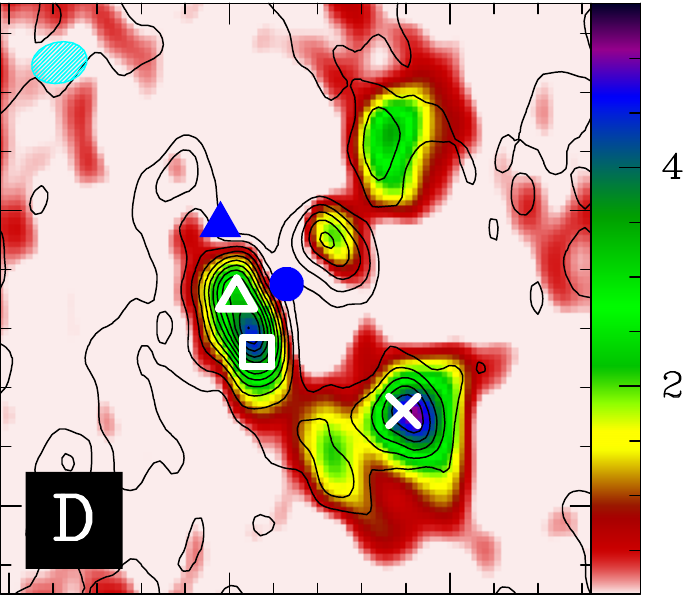}\vspace{0.2cm}\\
\includegraphics[scale=0.55,angle=0]{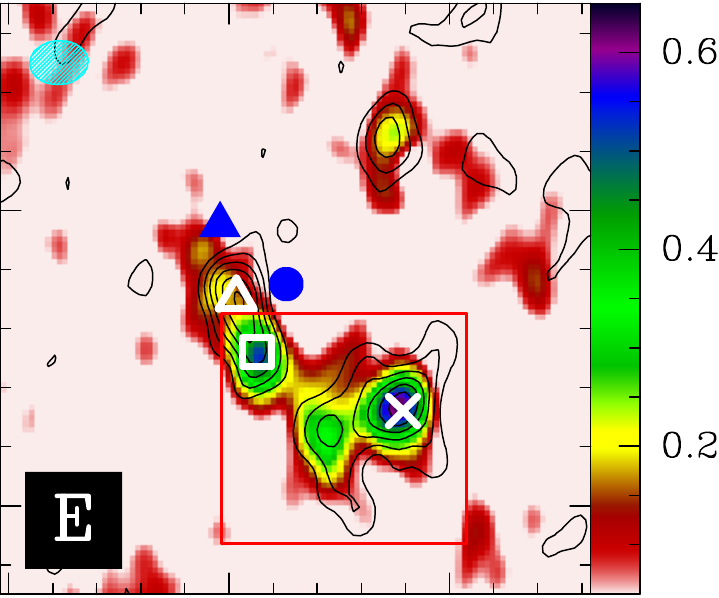}\vspace{0.13cm}\\
\includegraphics[scale=0.55,angle=0]{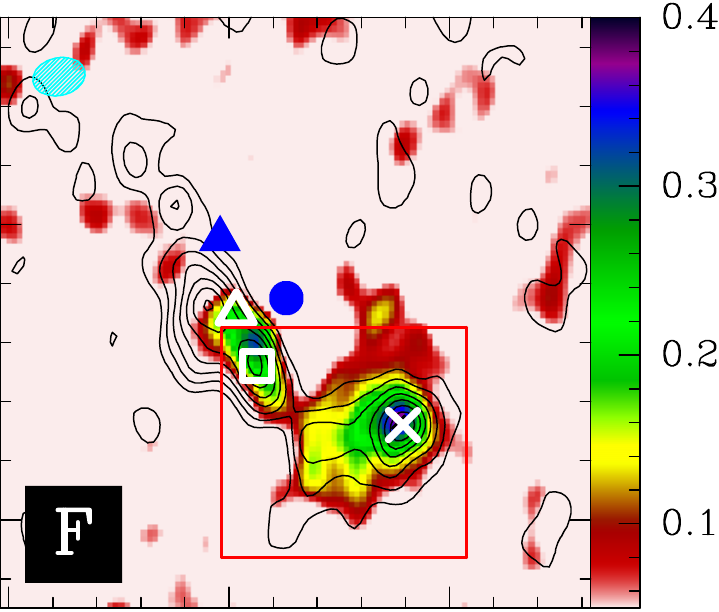}\vspace{0.2cm}\\
\hspace*{-1.13cm}\includegraphics[scale=0.55,angle=0]{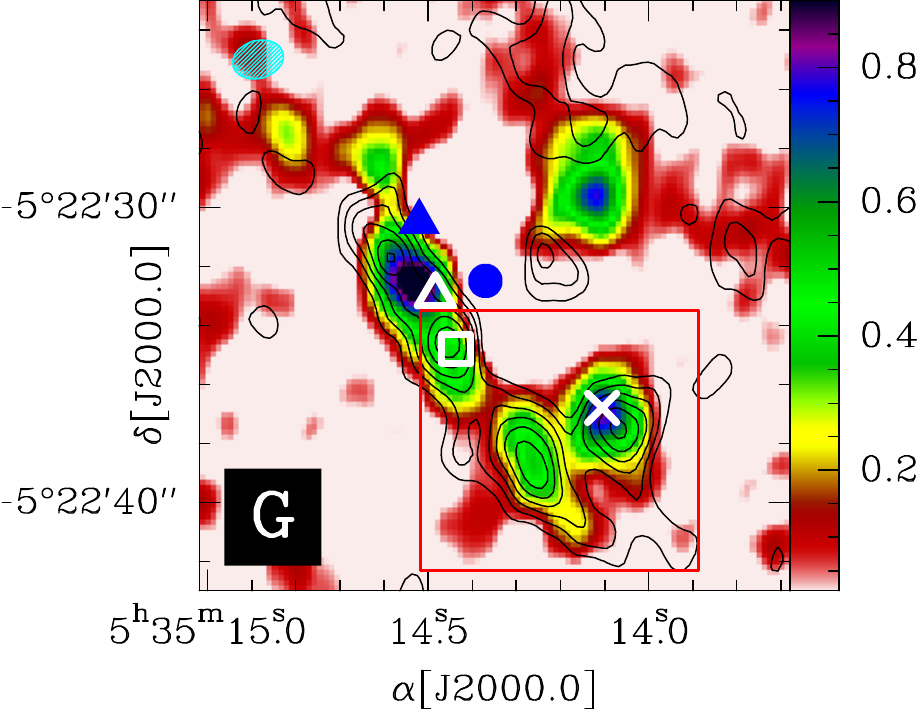}
\end{center}
\end{minipage}  
\begin{minipage}[t]{0.25\textwidth}
\begin{center}
\vspace*{-2.6cm}\includegraphics[scale=0.6,angle=0]{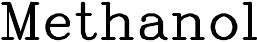}\\\vspace{0.2cm}
\includegraphics[scale=0.45,angle=0]{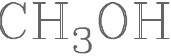}\\
\includegraphics[scale=0.45,angle=0]{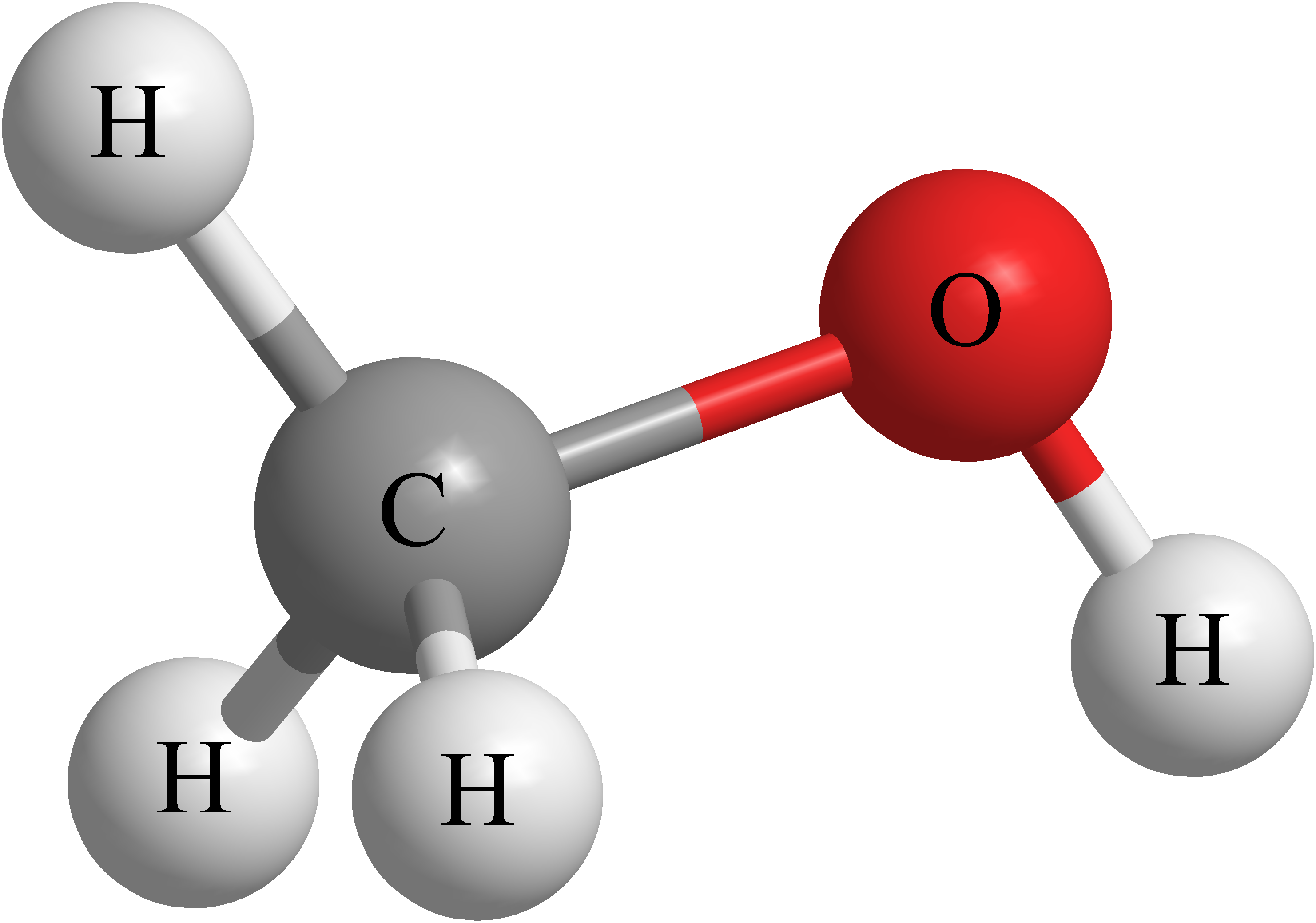}\\\vspace*{0.6cm}
\includegraphics[scale=0.6,angle=0]{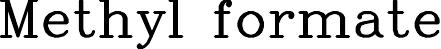}\\\vspace{0.2cm}
\includegraphics[scale=0.45,angle=0]{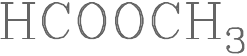}\\\vspace{0.2cm}
\includegraphics[scale=0.6,angle=0]{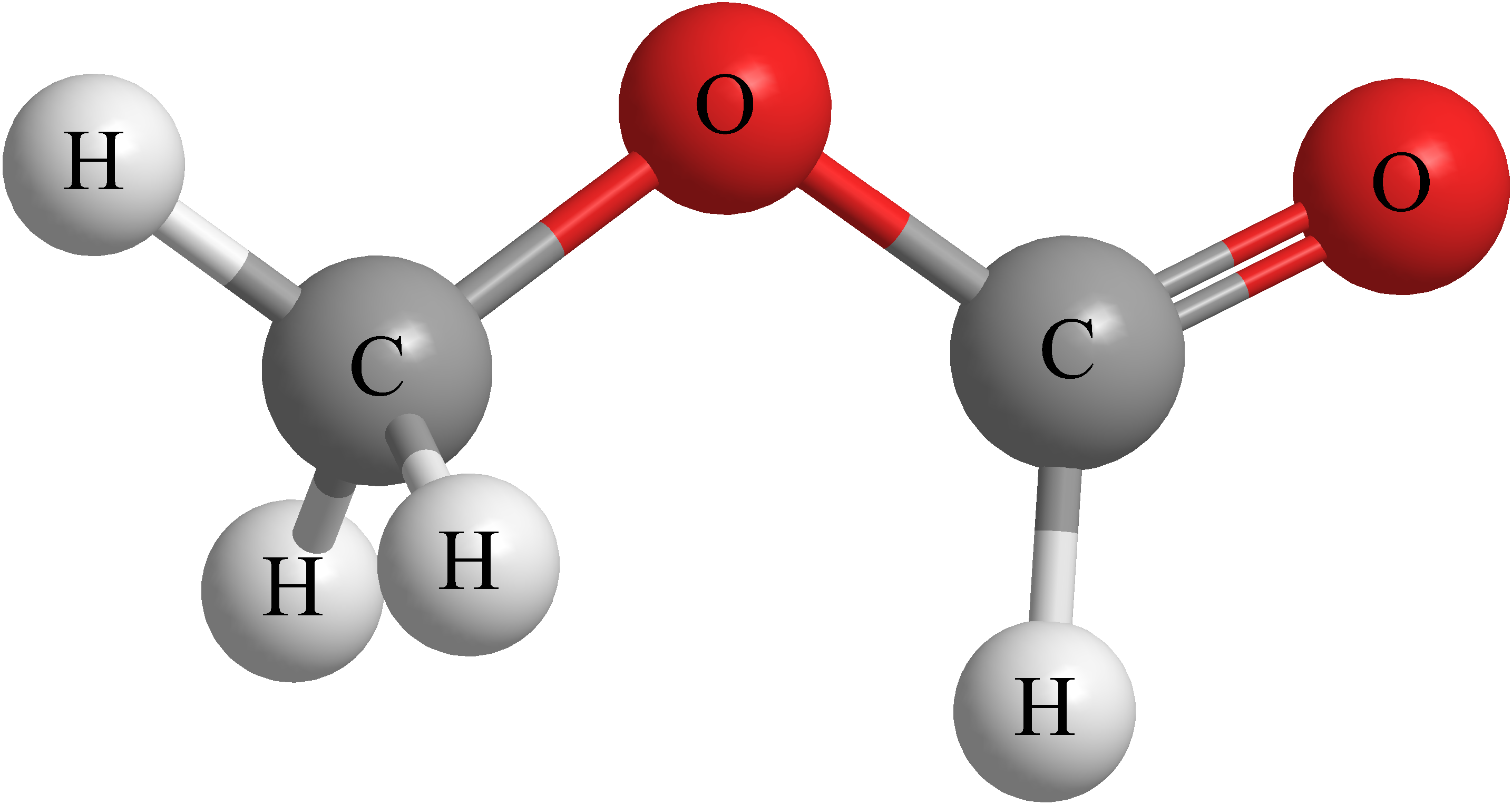}\\\vspace*{0.6cm}
\includegraphics[scale=0.6,angle=0]{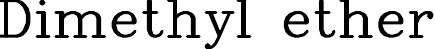}\\\vspace{0.2cm}
\includegraphics[scale=0.45,angle=0]{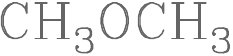}\\\vspace{0.2cm}
\includegraphics[scale=0.56,angle=0]{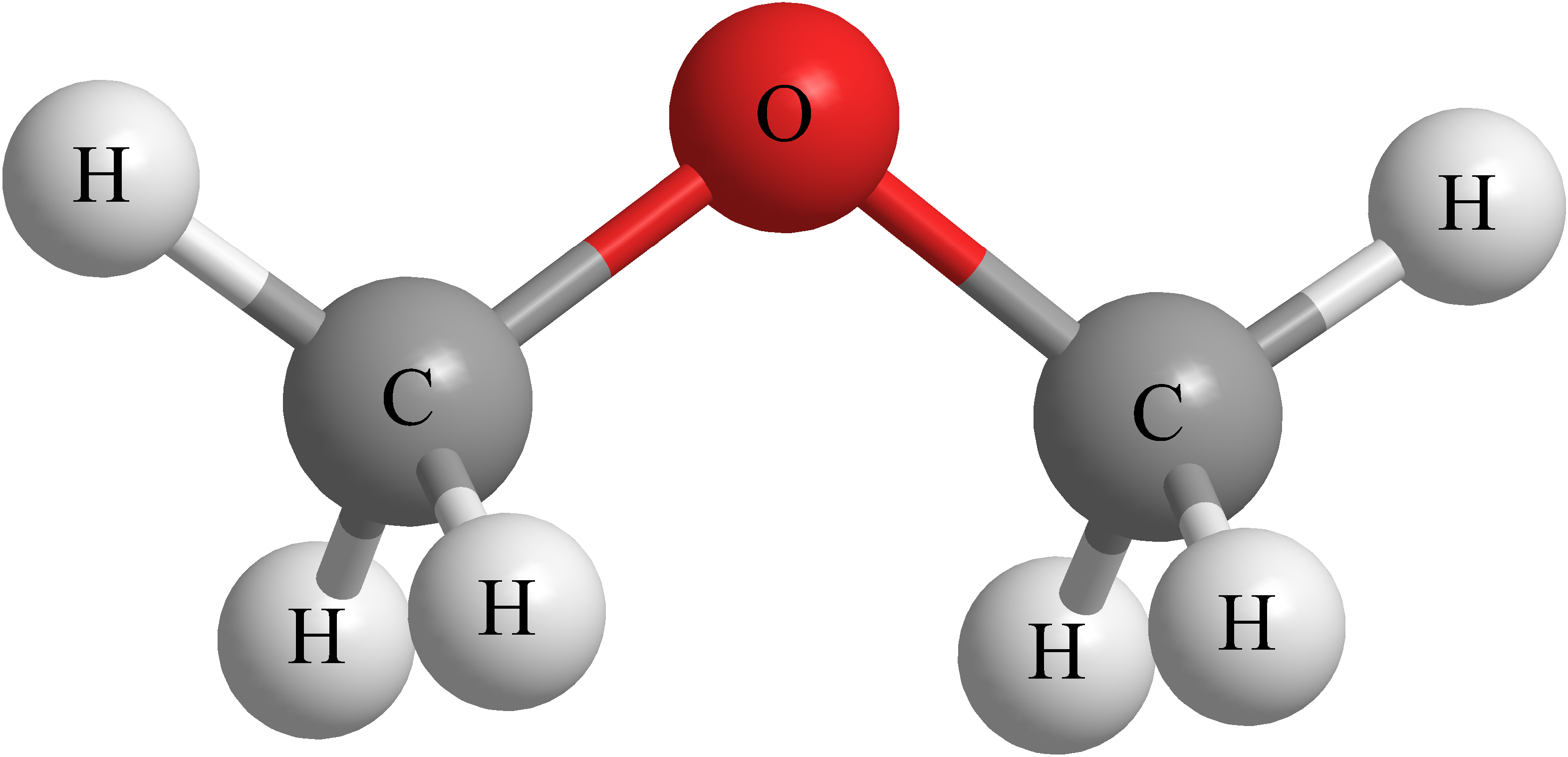}\\\vspace*{0.4cm}
\includegraphics[scale=0.6,angle=0]{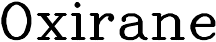}\\\vspace{0.2cm}
\includegraphics[scale=0.45,angle=0]{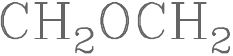}\\\vspace{0.25cm}
\includegraphics[scale=0.47,angle=0]{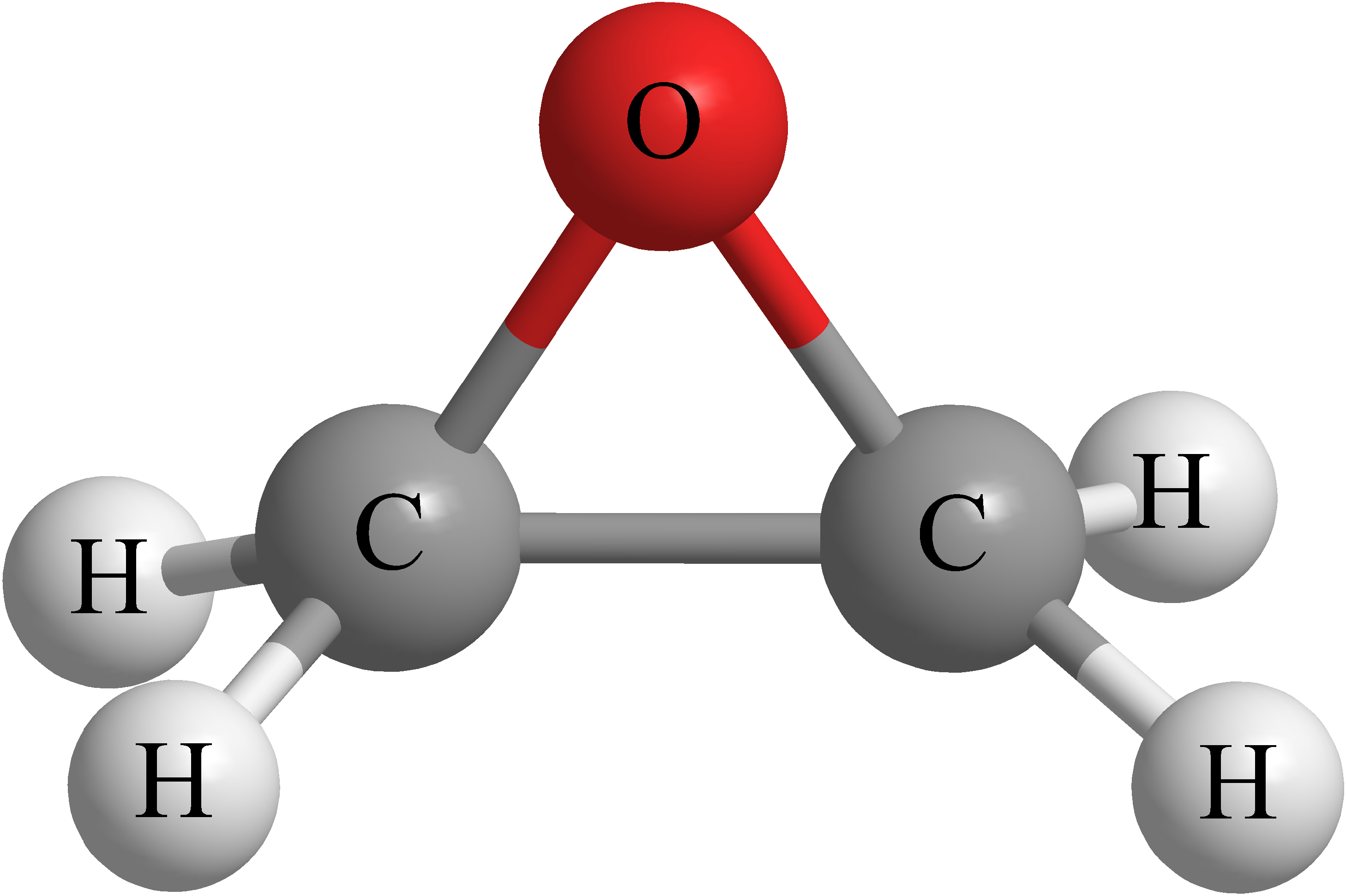}\\\vspace*{0.3cm}
\includegraphics[scale=0.6,angle=0]{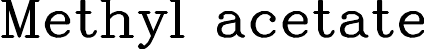}\\\vspace{0.2cm}
\includegraphics[scale=0.45,angle=0]{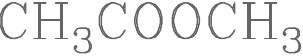}\\\vspace{-0.5cm}
\hspace*{0.5cm}\includegraphics[scale=0.6,angle=0]{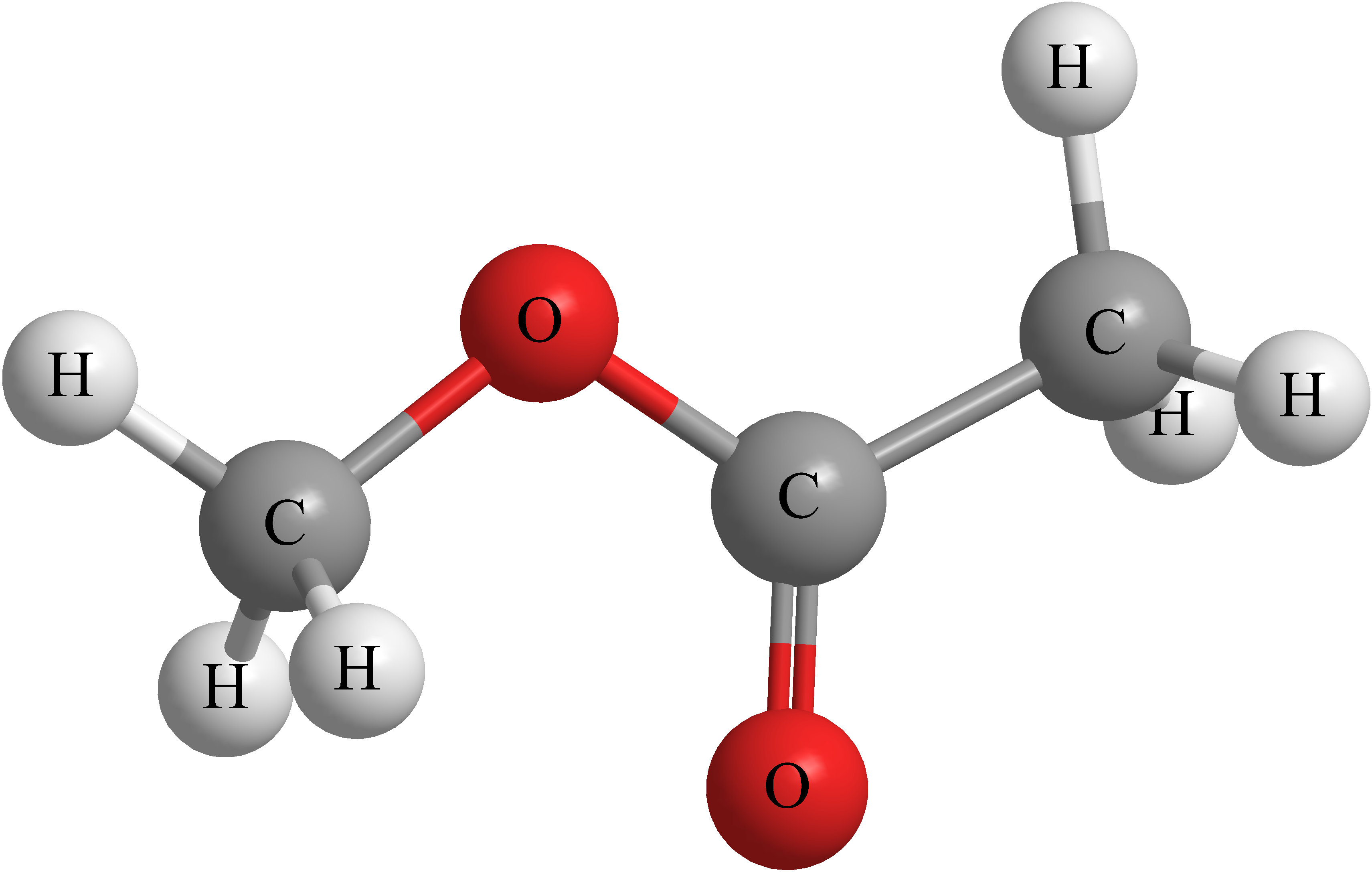}\\\vspace*{0.3cm}
\includegraphics[scale=0.6,angle=0]{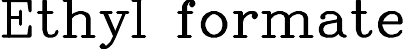}\\\vspace*{0.2cm}
\includegraphics[scale=0.45,angle=0]{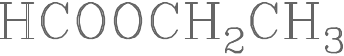}\\
\hspace*{0.5cm}\includegraphics[scale=0.65,angle=0]{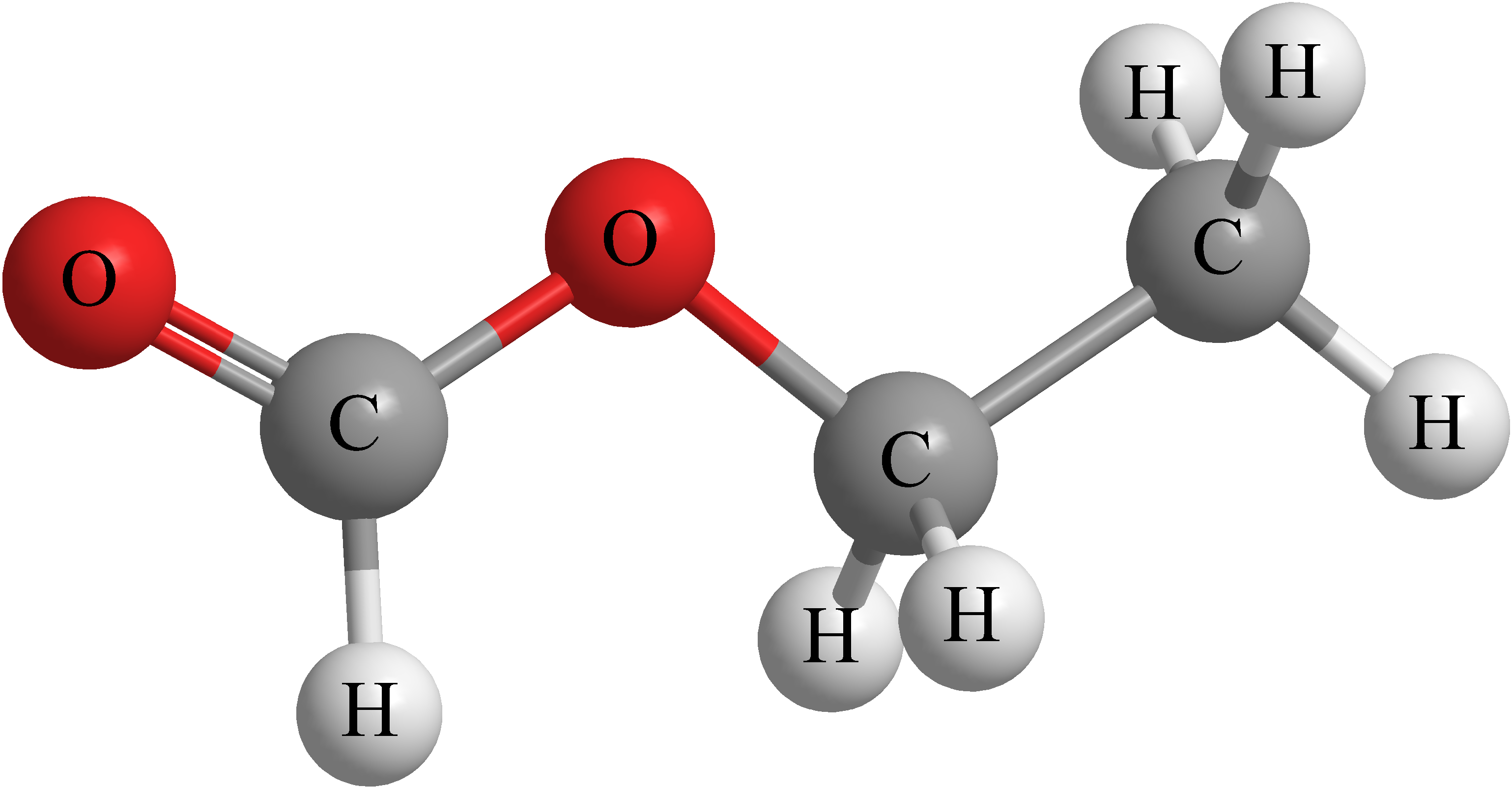}\\\vspace*{0.5cm}
\includegraphics[scale=0.6,angle=0]{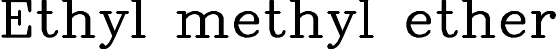}\\\vspace*{0.2cm}
\includegraphics[scale=0.45,angle=0]{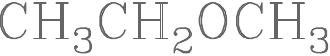}\\
\hspace*{0.4cm}\includegraphics[scale=0.6,angle=0]{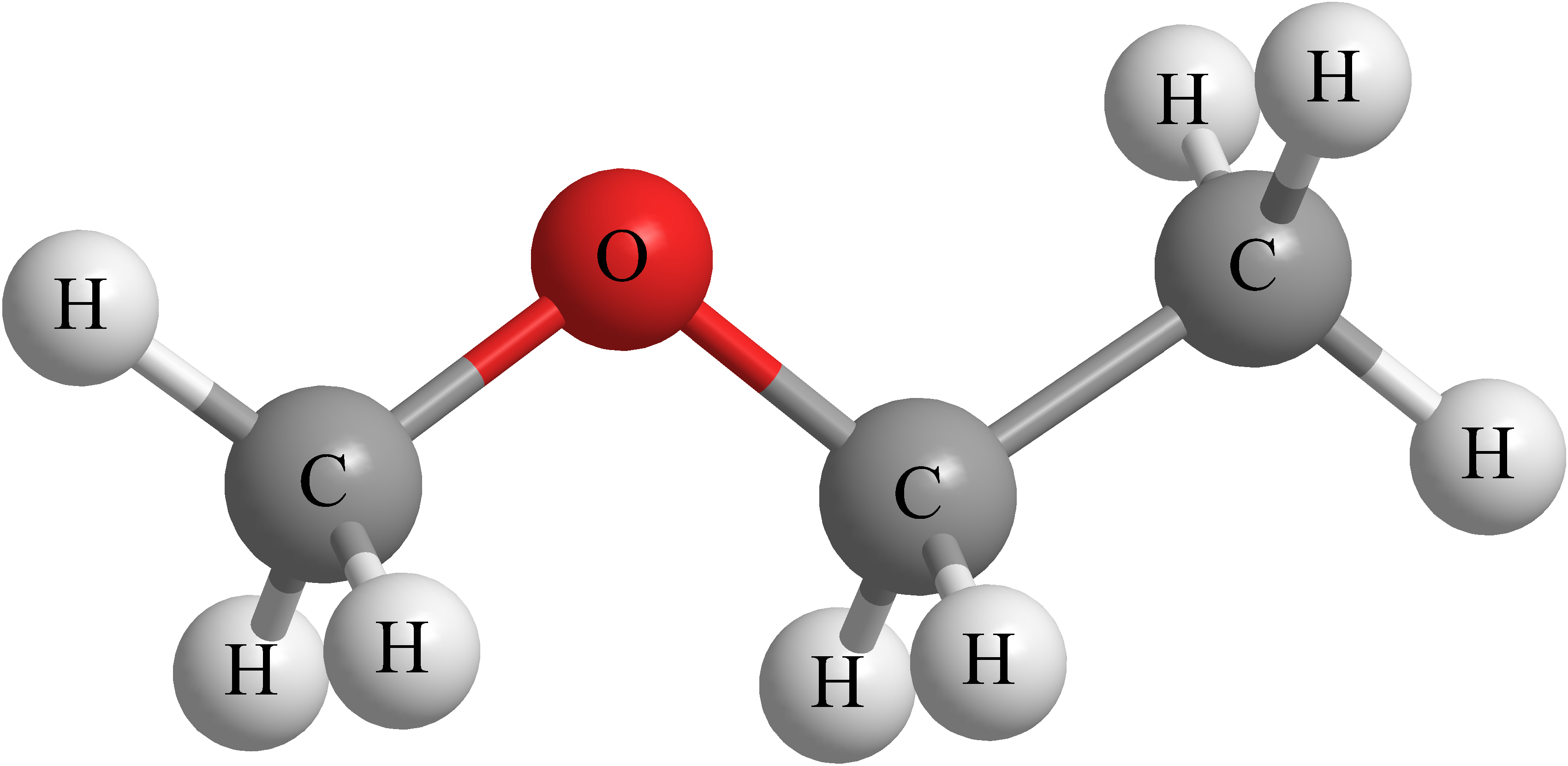}\\
\end{center}
\end{minipage} 
\begin{minipage}[t]{0.25\textwidth}
\begin{center}
\vspace*{-3cm}
\hspace*{1.5cm}\includegraphics[scale=0.55,angle=0]{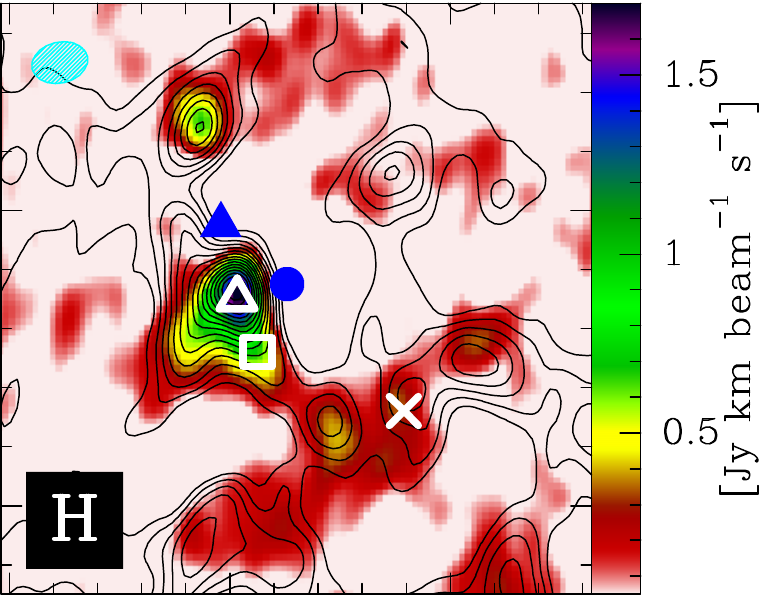}\vspace{0.2cm}\\
\hspace*{1.5cm}\includegraphics[scale=0.55,angle=0]{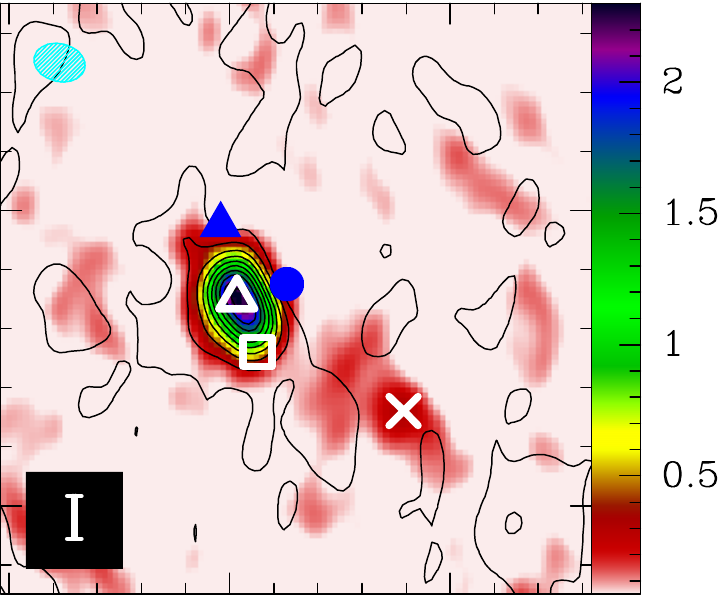}\vspace{0.2cm}\\
\hspace*{1.5cm}\includegraphics[scale=0.55,angle=0]{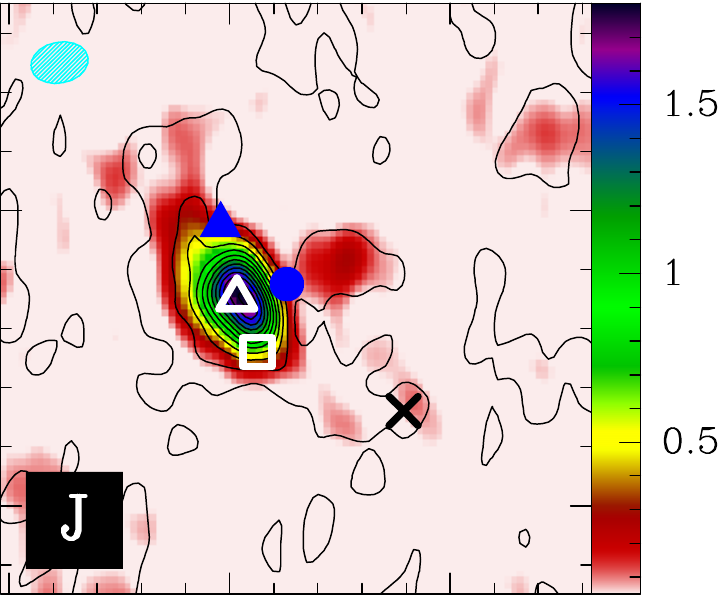}\vspace{0.2cm}\\
\hspace*{1.49cm}\includegraphics[scale=0.55,angle=0]{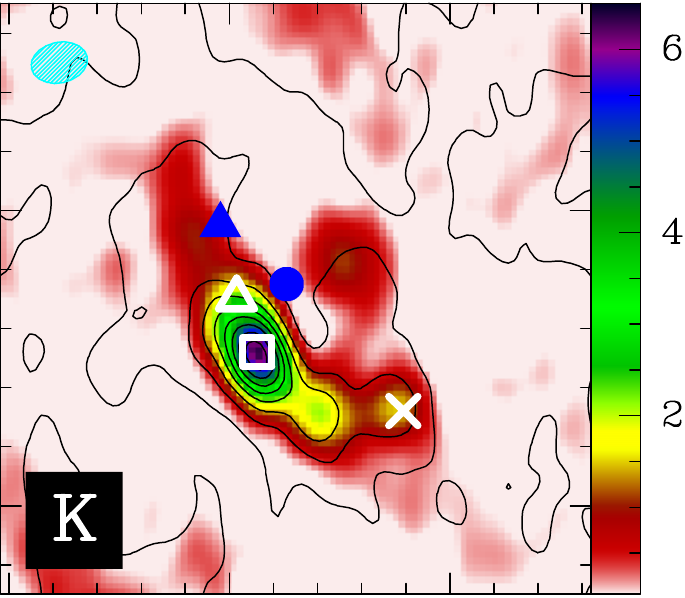}\vspace{0.13cm}\\\hspace*{1.43cm}
\includegraphics[scale=0.55,angle=0]{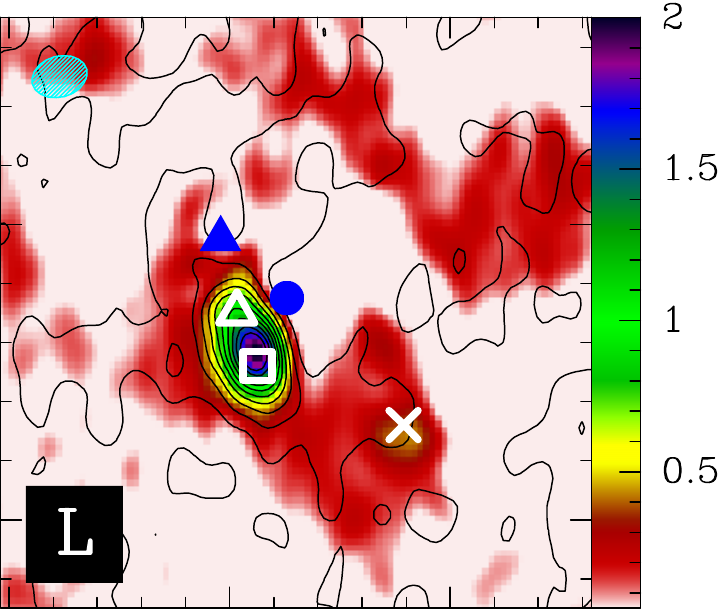}\vspace{0.2cm}\\\hspace*{1.43cm}
\includegraphics[scale=0.55,angle=0]{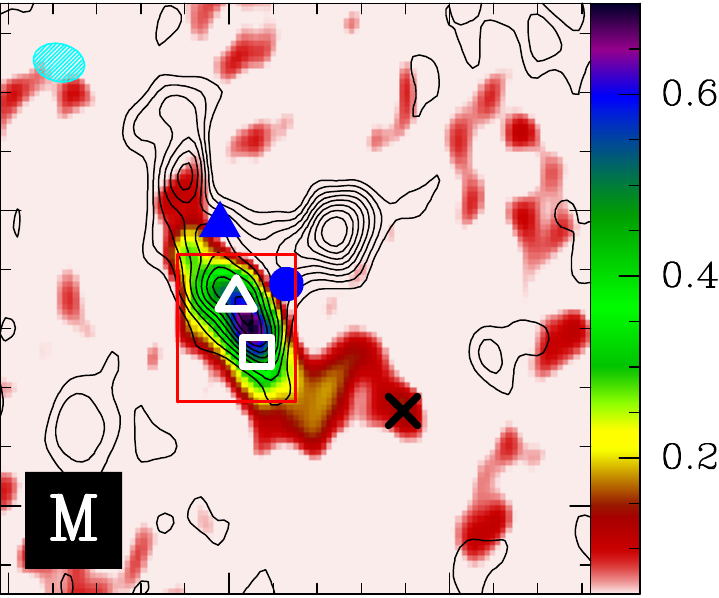}\vspace{0.2cm}\\
\hspace*{0.39cm}\includegraphics[scale=0.55,angle=0]{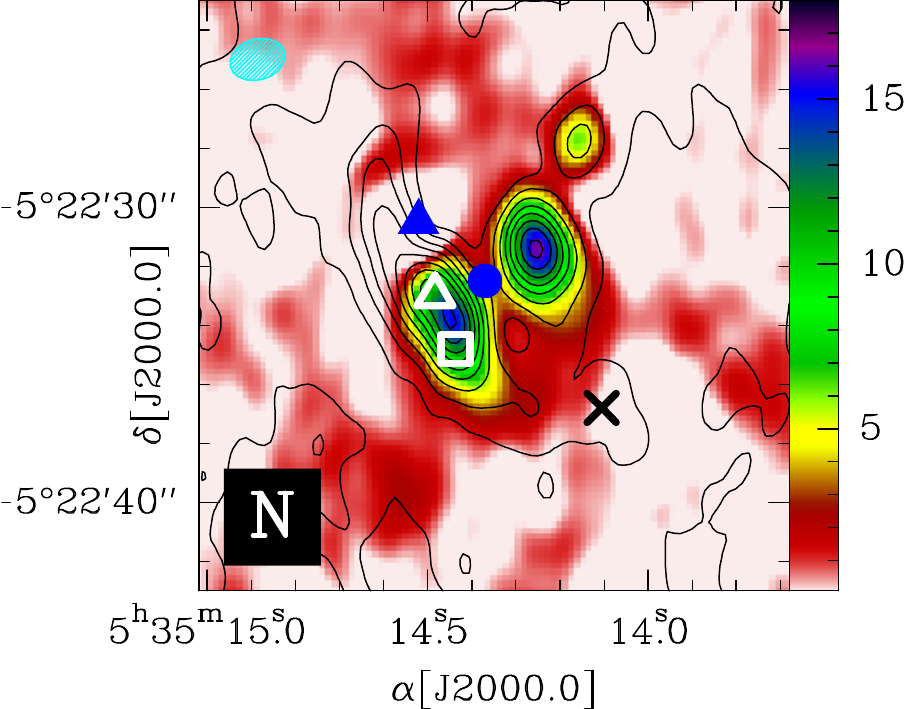}
\end{center}
\end{minipage} 
\begin{minipage}[t]{0.33\textwidth}
\begin{center}
\vspace*{-2.6cm}\includegraphics[scale=0.6,angle=0]{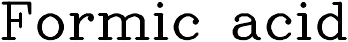}\\ \vspace{0.2cm}
\includegraphics[scale=0.45,angle=0]{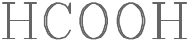}\\
\vspace*{-0.2cm}\hspace{-0.3cm}\includegraphics[scale=0.47,angle=0]{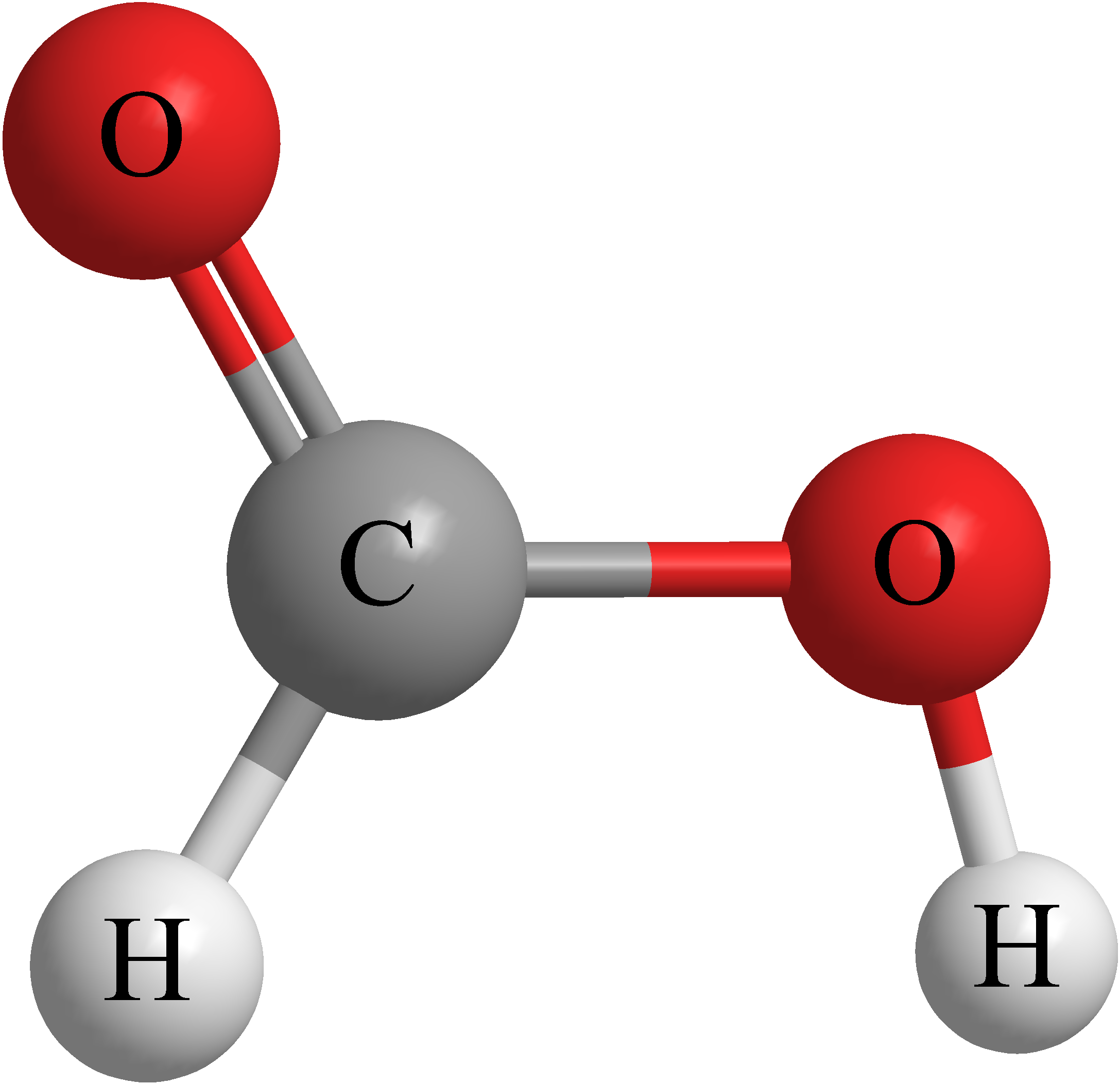}\\
\vspace*{0.4cm}
\includegraphics[scale=0.6,angle=0]{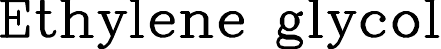}\\\vspace*{0.2cm}
\includegraphics[scale=0.45,angle=0]{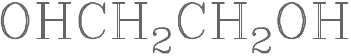}\\\vspace*{0.2cm}
\hspace{0.5cm}\includegraphics[scale=0.5,angle=0]{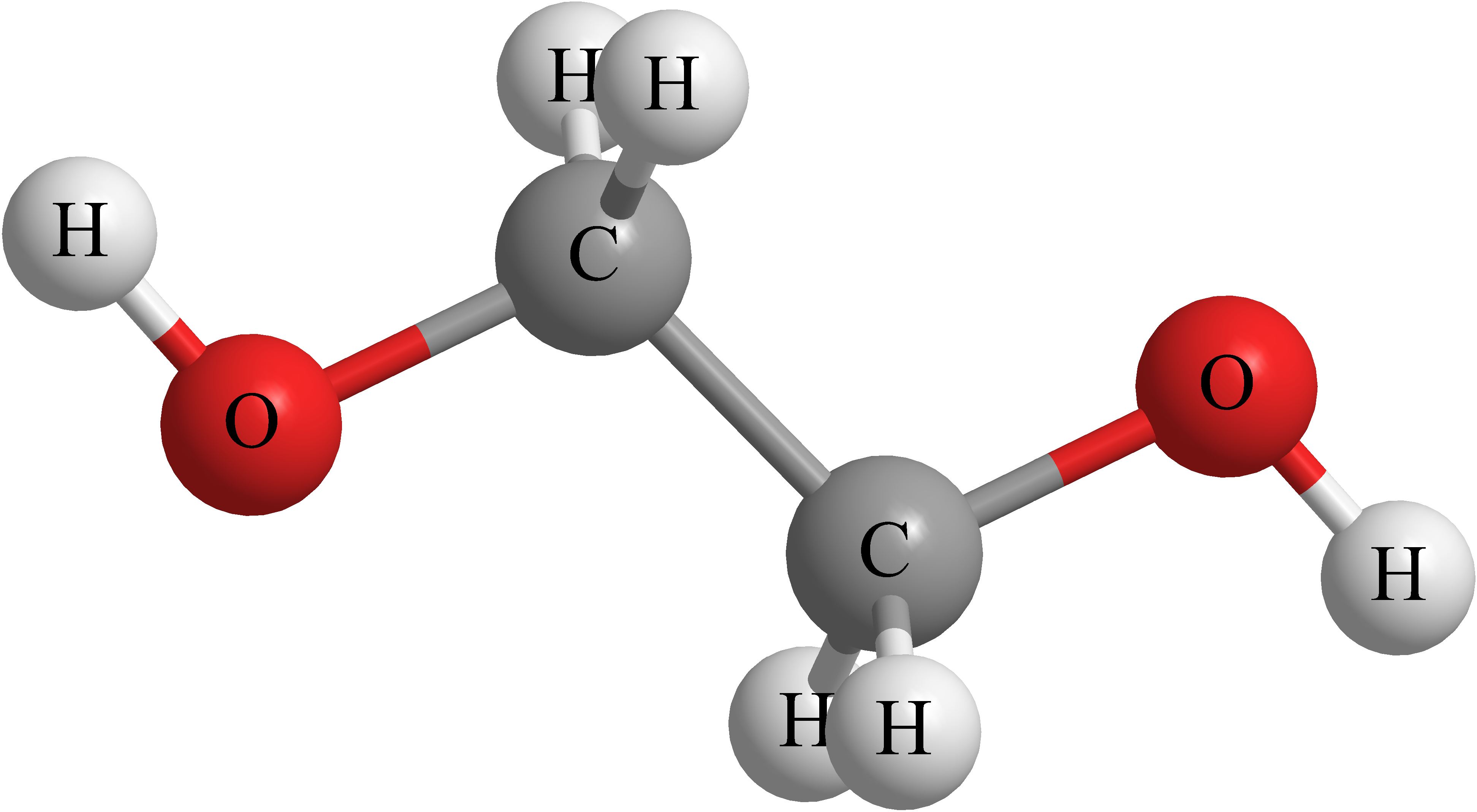}\\\vspace*{0.25cm}
\includegraphics[scale=0.6,angle=0]{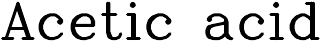}\\\vspace*{0.2cm}
\includegraphics[scale=0.45,angle=0]{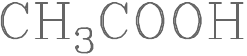}\\\vspace*{-0.2cm}
\hspace*{1.2cm}\includegraphics[scale=0.45,angle=0]{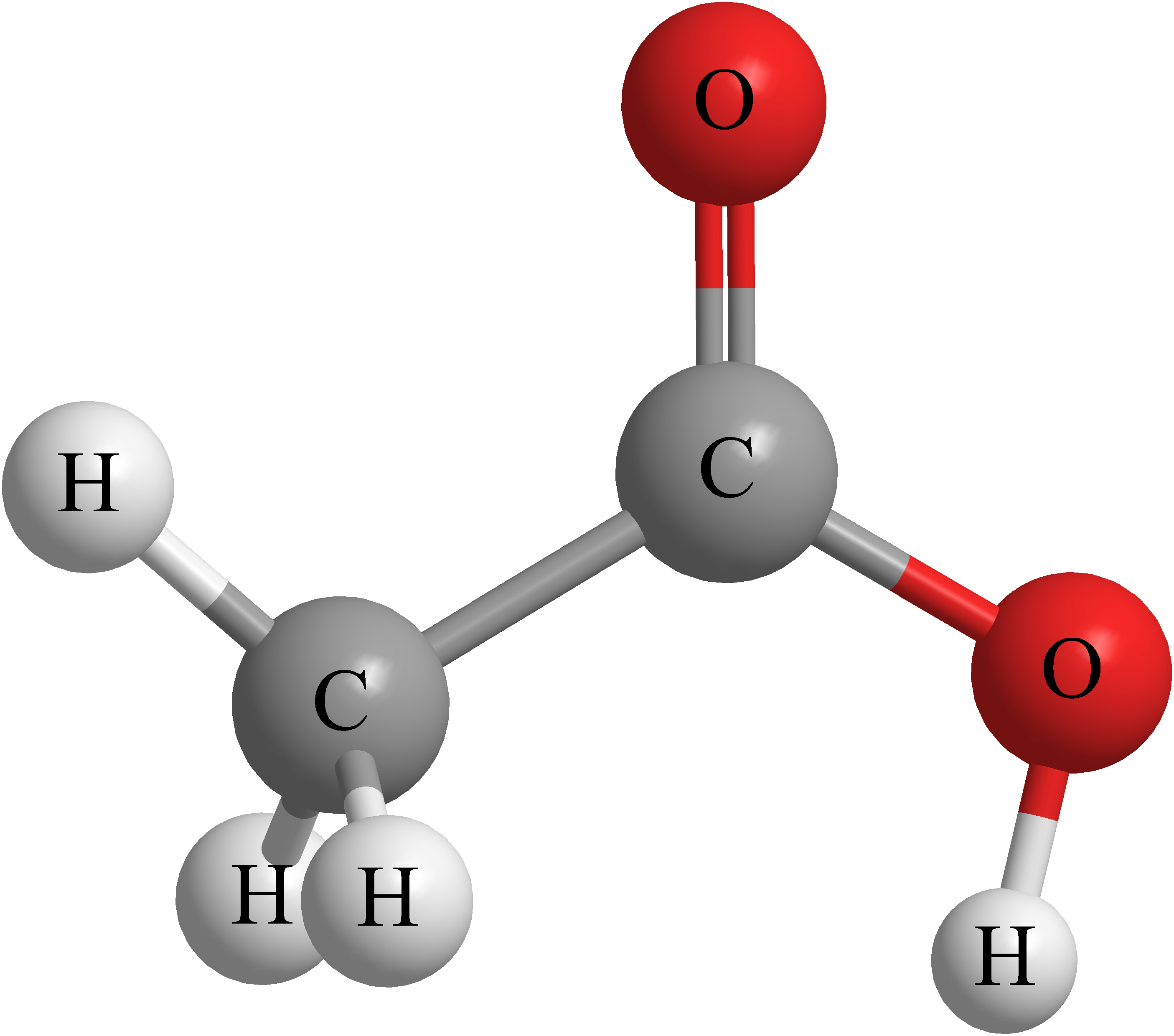}\\\vspace*{0.3cm}
\includegraphics[scale=0.6,angle=0]{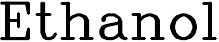}\\\vspace*{0.2cm}
\includegraphics[scale=0.45,angle=0]{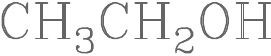}\\\vspace*{-0.3cm}
\hspace*{1cm}\includegraphics[scale=0.63,angle=0]{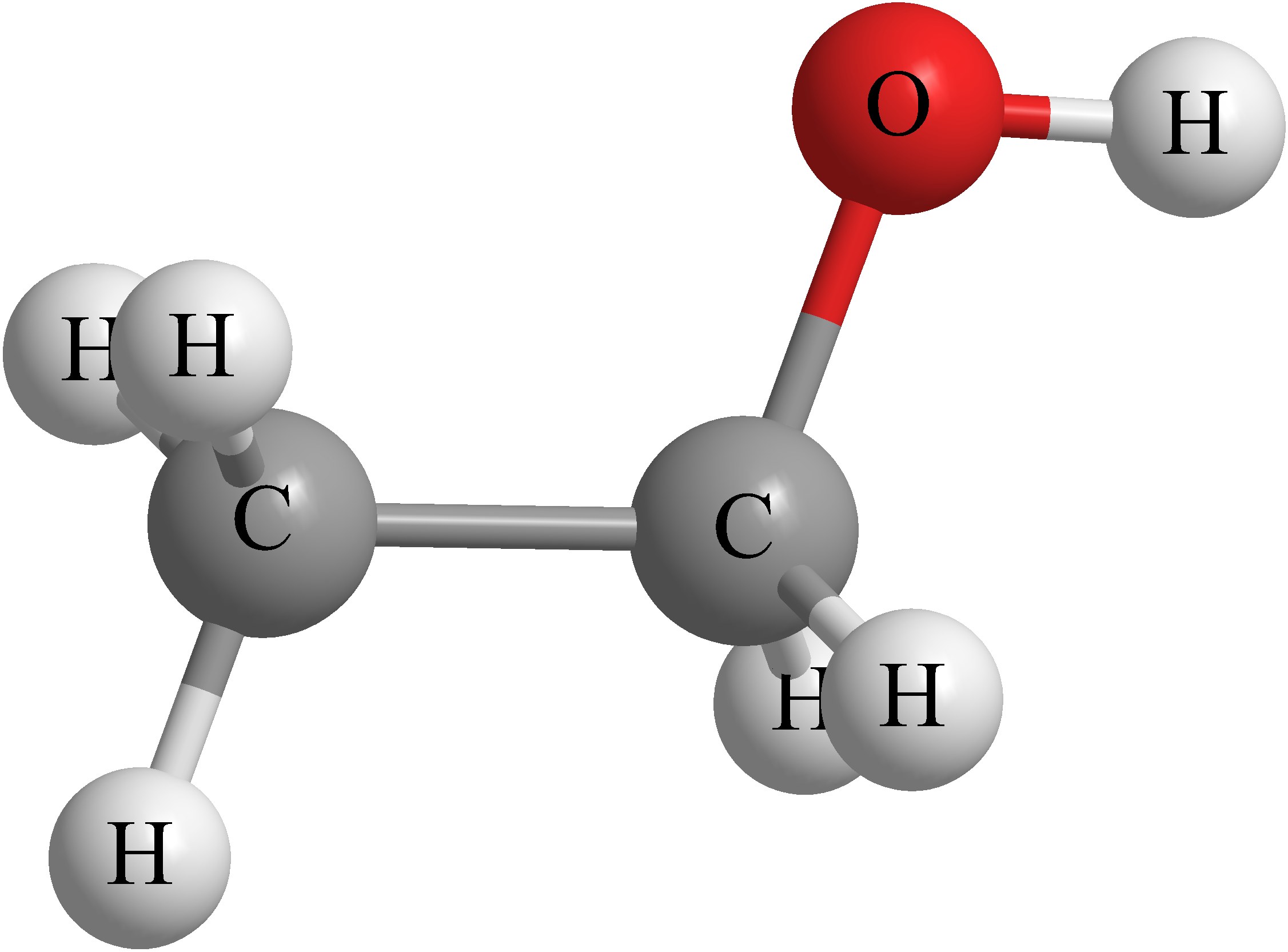}\\\vspace*{0.6cm}
\includegraphics[scale=0.6,angle=0]{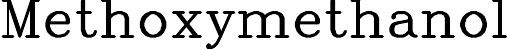}\\\vspace*{0.2cm}
\includegraphics[scale=0.45,angle=0]{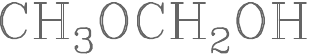}\\\vspace*{0.2cm}
\includegraphics[scale=0.55,angle=0]{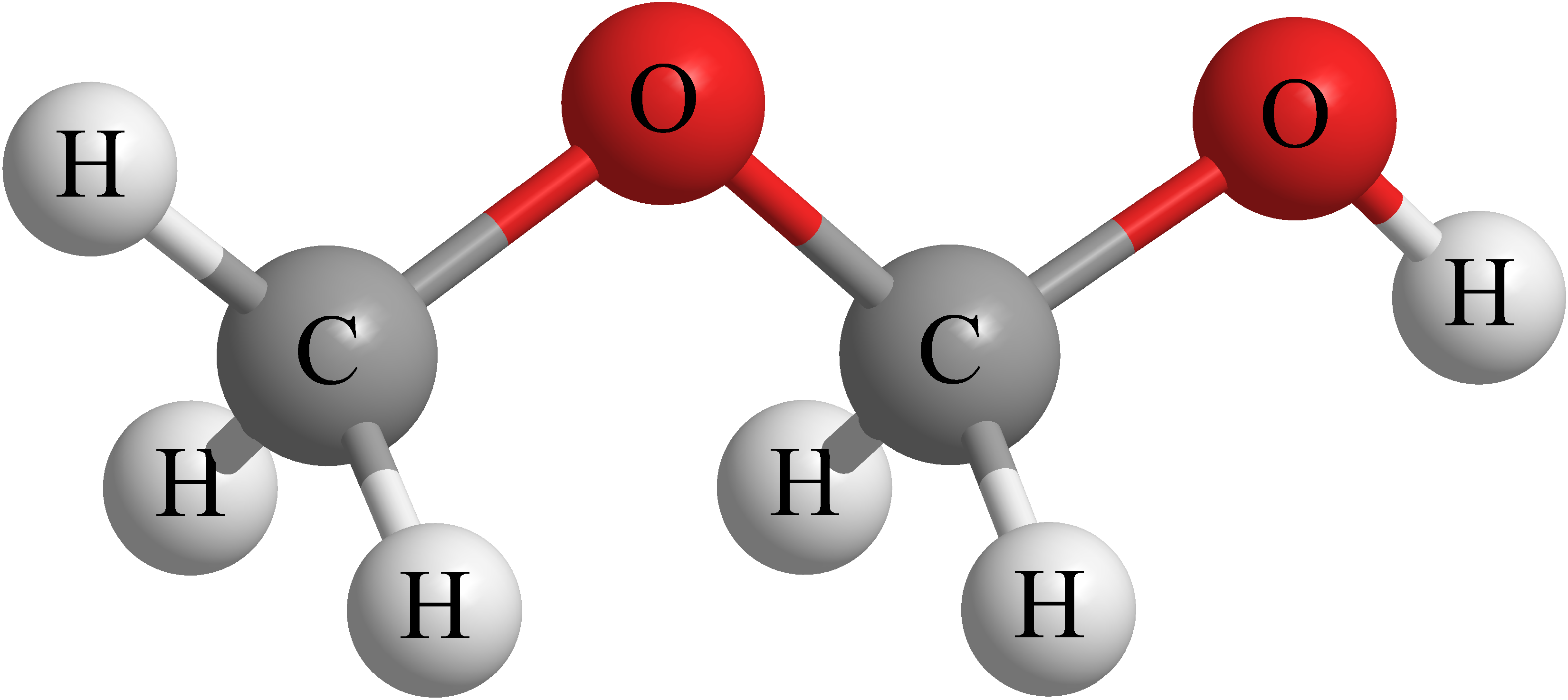}\\\vspace*{0.7cm}
\includegraphics[scale=0.6,angle=0]{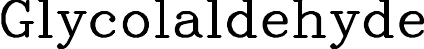}\\\vspace*{0.2cm}
\includegraphics[scale=0.45,angle=0]{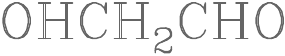}\\\vspace*{-0.1cm}
\includegraphics[scale=0.45,angle=0]{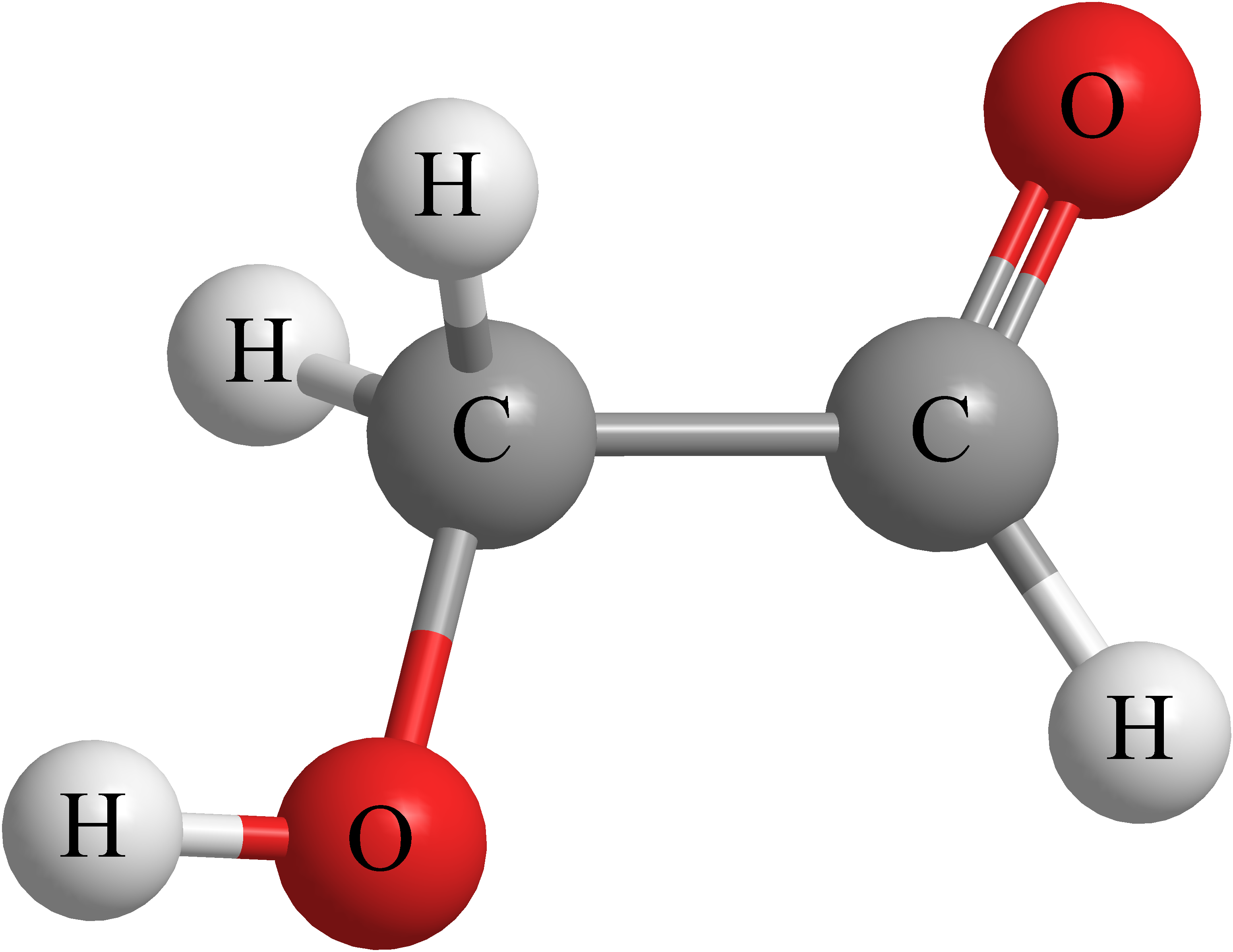}\\\vspace*{0.4cm}
\includegraphics[scale=0.6,angle=0]{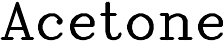}\\\vspace*{0.2cm}
\includegraphics[scale=0.45,angle=0]{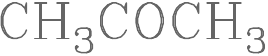}\\\vspace*{0.15cm}
\includegraphics[scale=0.47,angle=0]{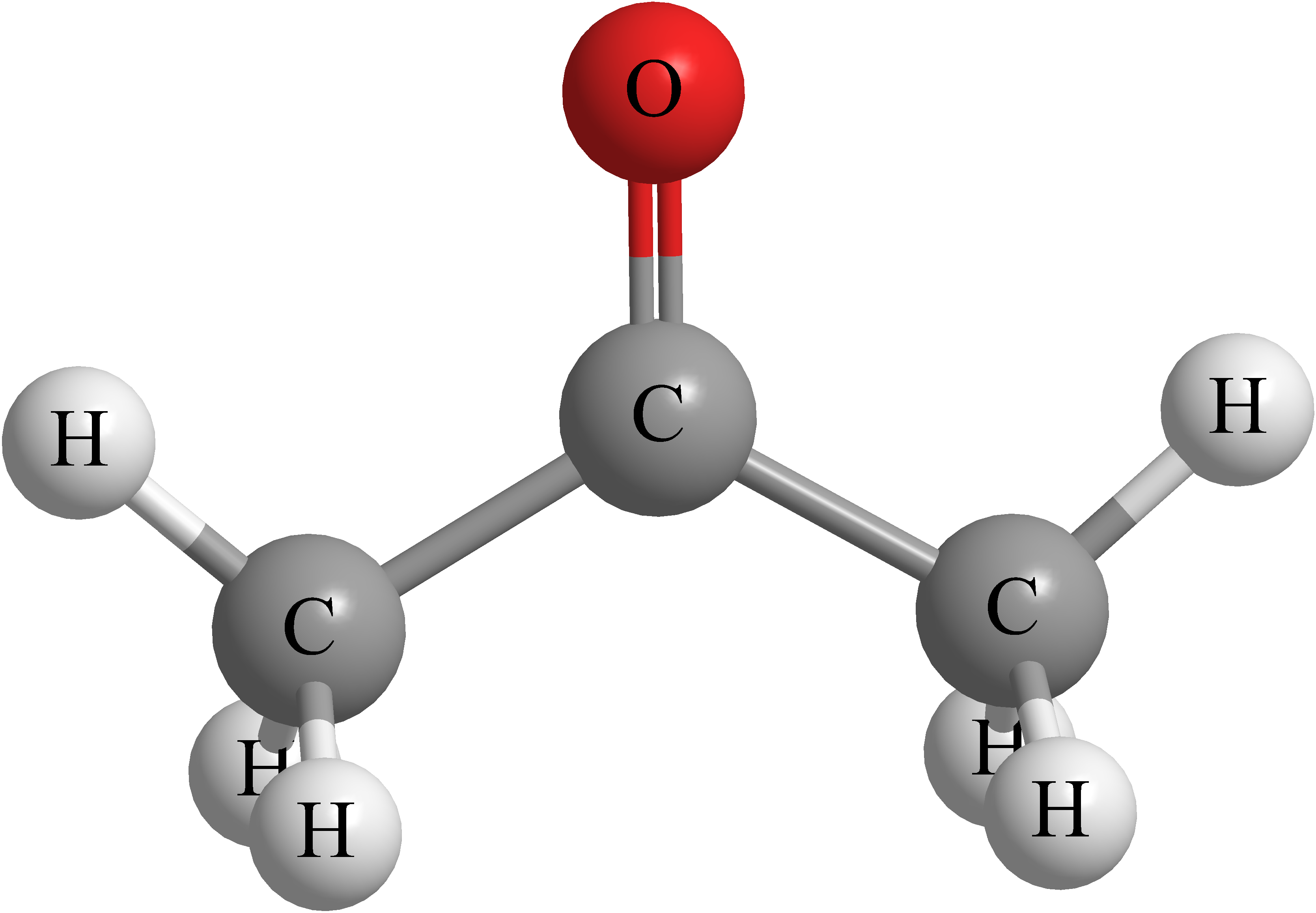}\\
\end{center}
\end{minipage} 
\caption{Spatial distribution of O-bearing species in Orion KL (see text, Sect.\,\ref{results}).
The different positions discussed in the text are indicated by symbols (blue triangle: source $I$; blue circle: source $n$;
cross: MF peak; unfilled triangle: EG peak; unfilled square: ET peak).
The cyan ellipse in the top left corner
of each map represents the ALMA synthetic beam. Red rectangles confine the assumed spatial distribution for E, F, G, and M species. The values of the levels are shown in Table\,\ref{table_lines}.}
\label{fig_maps}
\end{figure*}

\section{Results}\label{results}

Several O-bearing COMs have been identified towards different Orion components.
In this work we study the most complex ones together with some related species:
alcohols, $^{13}$CH$_3$OH, CH$_3$CH$_2$OH, OHCH$_2$CH$_2$OH, and CH$_3$OCH$_2$OH; ethers, CH$_3$OCH$_3$, CH$_2$OCH$_2$, and CH$_3$CH$_2$OCH$_3$;
ketones, CH$_3$COCH$_3$; aldehydes, OHCH$_2$CHO (only tentatively detected);
esters, HCOOCH$_3$, HCOOCH$_2$CH$_3$, and CH$_3$COOCH$_3$; and carboxylic acids, HCOOH and CH$_3$COOH.
We have detected methoxymethanol for the first time in Orion (for laboratory spectroscopy of methoxymethanol see \citealt{Motiyenko_2018}) and the second in interstellar medium
\citep{McGuire2017}.

Owing to the high spatial resolution of the ALMA SV data that distinguishes between the contributions from the
different cores of Orion, we constrained the spatial distributions of these species and their abundances.
To perform the ALMA maps, we selected two transitions of each of these species that arise free of blending with other species in
the Orion KL cores (see Table\,\ref{table_lines} and Fig.\,\ref{fig_lines_1}). 
When possible, the two selected transitions have large differences in the energy of the upper level state.
Then, for each species, the spatial distribution of the integrated line emission of the two 
selected lines was overlapped by depicting it in different visual scales (colour and contours; see Fig.\,\ref{fig_maps}).
Doing so, we addressed possible effects in the spatial distribution related to
different excitation conditions and
potential blendings of the lines with other species in the different components.

Moreover, to limit opacity effects in the emission distribution of the most abundant species,
we mapped $a$-type transitions of $^{13}$CH$_3$OH and $b$-type transitions of HCOOCH$_3$
(those associated with the lowest dipole moment of each species).
Despite the lack of short spacing data required for completion of the possible extended emission
of some of these species (see e.g. $^{13}$CH$_3$OH or HCOOH), we obtain an overview of the main cores that host the
studied species.

We note the coincidence in the emission peaks between colour and contour scales for most of the species (see Fig.\,\ref{fig_maps}). Only for the less abundant species
are there some regions which exhibit emission only in one of the depicted lines due to light blendings with other molecular species in a particular
region of the cloud (see Table\,\ref{table_lines} and Fig.\,\ref{fig_lines_1}). In those cases (CH$_3$COOCH$_3$,
HCOOCH$_2$CH$_3$, CH$_3$CH$_2$OCH$_3$, and OHCH$_2$CHO) we assumed only the overlapping areas (colour\,+\,contours) as the emitting regions
for these species (emission confined inside red rectangles in Fig.\,\ref{fig_maps}).

Figure\,\ref{fig_maps} shows the ALMA maps together with
the molecular structure of the corresponding species for all complex O-bearing molecules mentioned above.
At first glance there is a clear differentiation in the spatial distribution of the species
situated at the two different sides of Fig.\,\ref{fig_maps}: the molecules on the left (panels A--G) present the bulk of the emission at the position of the compact ridge (crosses), the emission of the species on the right (panels H-N)
is characterised by a lack of a significant contribution towards this component.
Moreover, the species on the right do not present a common emission peak towards their main emitting region (the hot core
south). We distinguished two locations separated by about 
$\sim$1.6$''$ for the                                                                                                                                                                                                                                                                                                                                                                                                                                                                                                                                                                                                                                                                                                                                                                                                                                                                                                                                                                                                                                                                                                                                                                                                                                                                                                                                                                                              
main emission peaks in the hot core south: the maximum of formic acid (H), ethylene glycol (I), and acetic acid (J), in agreement with the position found
in other works \citep{Brouillet_2015,Cernicharo_2016,Favre_2017,Pagani_2017} for these species (unfilled triangles)
and the maximum of ethanol (K) and methoxymethanol (L) (unfilled squares).
In addition, we note that the species which present compact ridge emission (A--G)
also exhibit  emission in the hot core south showing a maximum in a position near the unfilled square (emission peak of ethanol and methoxymethanol).
It is worth noting that a clump at the north-west appears in the emission of species A, B, C, D, and E, 
thus presenting the chemical fingerprints found for the compact ridge (see e.g. \citealt{Favre_2011a}, \citealt{Wu_2014}, 
and \citealt{Hirota_2015} for further discussion
of this component). However, we cannot confirm 
emission from the less abundant species of this group (F and G) towards this component.

Owing to the observed spatial distributions of these species, we distinguish three emission peaks for further discussions (see Fig.\,\ref{fig_maps}):
MF (methyl formate peak, cross symbol), EG (ethylene glycol peak, unfilled triangle), and ET (ethanol peak, unfilled square).
Coordinates for these positions are shown in Table\,\ref{table_cd}. Interestingly, these emission peaks coincide with three continuum sources found
by \citet{Hirota_2015} at 245\,GHz and 339\,GHz using subarcsecond resolution observations performed with the ALMA interferometer:
HKKH~11 for the MF~peak, HKKH~9 for the ET~peak, whereas HKKH~8 is close to EG~peak
(see Fig.\,4 and Table\,3 of \citealt{Hirota_2015}). In addition, \citet{Wright_2017} show that these continuum
sources lie along the boundaries of the SiO outflow emission, being HKKH~8 (EG~peak) the more embedded source 
in the SiO $J$\,=\,2\,$\rightarrow$\,1 emission (see Fig.\,1 of \citealt{Wright_2017}). Moreover,
\citet{Brouillet_2015} noted that one short CO jet identified by \citet{Zapata_2009}
overlaps the EG~peak, which suggests that this component may also be located in the path of the high-velocity outflow.

We used the MADEX\footnote{https://nanocosmos.iff.csic.es/?page$\_$id=1619} 
radiative transfer code \citep{Cernicharo_2012} to derive physical parameters and column densities for the studied species 
towards the MF, EG, and ET~peaks (see Appendix\,\ref{appendix_b}).
Figure\,\ref{fig_lines_1} shows the lines mapped in Fig.\,\ref{fig_maps} (colour and contours) in the three selected positions
(MF, EG, and ET peaks) together with the adopted integrated area for imaging the spatial distribution of these lines and
the synthetic spectrum provided by MADEX. In addition, to ensure the first detection of methoxymethanol in Orion, 
Fig.\,\ref{fig_ch3och2oh} shows selected lines of CH$_3$OCH$_2$OH in the ET peak together with the model 
derived using MADEX according to the physical parameters given in Table\,\ref{table_cd}.

\begin{figure}[!ht]
\includegraphics[scale=0.355,angle=0]{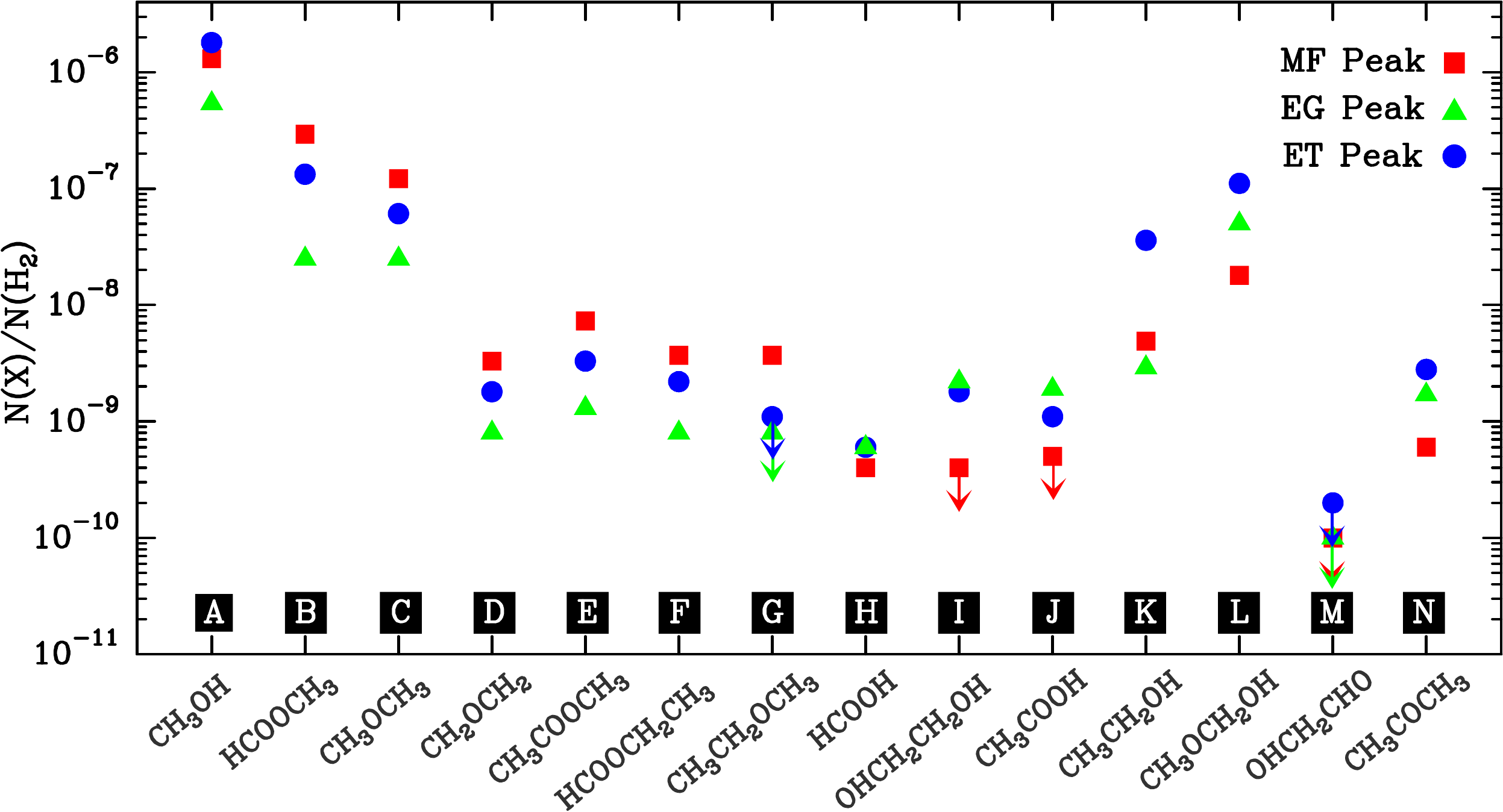}
\caption{Abundances of complex O-bearing species in Orion KL.}
\label{fig_abun}
 \end{figure}

Molecular abundances for each species in the different positions were calculated using
the derived column densities shown in Table\,\ref{table_cd}.  
We adopted the $N$(H$_2$) values of~8.2\,$\times$\,10$^{23}$, 2.4\,$\times$\,10$^{24}$, and 
1.8\,$\times$\,10$^{24}$~cm$^{-2}$ derived by \citet{Hirota_2015} in HKKH~11~(MF), HKKH~8~(EG), and HKKH~9~(ET), respectively,
using a dust temperature of 100\,K and a modified blackbody fitting of the spectral energy distribution (SED).
Figure\,\ref{fig_abun} shows the molecular abundances obtained for the studied species in the different positions. Interestingly,
the derived values reproduce the same tendency observed in the maps. While species from A to G exhibit the
lowest abundance towards the EG~peak, the MF~peak is the position where species from H to N present lower abundances.

\section{Discussion}
\label{dis}

The clear spatial segregation of the studied O-bearing species
allows us to infer some common evident molecular features for these molecules
depending on whether they present significant emission in the compact ridge component or not.
All molecules from B to G harbour an oxygen bound to two carbons by two single bonds (C$-$O$-$C; an ether group),
whereas all species from H to M show a hydroxyl group ($-$OH) bound to a carbon atom (C$-$O$-$H).
This points to a principal common precursor for the formation (via dust and/or gas-phase reactions) of these species.
According to this molecular structure, we  propose methoxy (CH$_3$O$-$)
and hydroxymethyl ($-$CH$_2$OH) radicals as the major drivers of the chemistry
in the compact ridge and the hot core south, respectively.
Nevertheless, studies on laboratory experiments on ices and surface and/or gas-phase chemical models
could point in alternative directions.
Paying attention to these studies, the chemical fingerprints found in this work, the positions of MF, EG, and ET peaks along Orion KL,
and assuming similar initial conditions for the 
parent cloud, we could propose several scenarios for
the chemical segregation found in this paper.

\subsection{Surface and gas-phase chemistry}
Our results show that C$-$O$-$C-bearing species are present both at the hot core south and the compact ridge,
but peak at the latter.
Due to the recent explosive event (500$-$1000 yr) in Orion, we could expect a young stage for the 
chemistry of this region. The simplest model to explain the rather strong correlation
of these species is radical-radical reactions on grain surfaces (see \citealt{Garrod_2008}).
Then the relative abundances of the C$-$O$-$C-bearing species may be fixed
before this event and the absolute abundances could be enhanced after the explosion. 
Moreover, the hot core south
could be more exposed to the effects of this event (i.e. higher temperature, particles, or shocks) giving rise
to a surface chemistry which produces C$-$O$-$H-containing molecules. In this case, the mobility
enhancement of the CH$_2$OH radical could be the crucial phenomenon.

Nonetheless, although we have assumed a young chemistry for this source, 
the enhancement of the C$-$O$-$C-bearing species towards the compact ridge
together with lack of C$-$O$-$H-containing molecules in this component could point  alternatively to an important 
role of gas-phase reactions in this component.

Laboratory experiments on ices and surface chemical models
may suggest that both C$-$O$-$C- and C$-$O$-$H-containing molecules are released from grains to the gas phase
during the warm-up timescales:
\vspace{-\topsep}
\begin{itemize}
\item In laboratory experiments to quantify the production rates 
of COMs, \citet{Oberg_2009} 
identified CO, CO$_2$, CH$_4$, HCO, 
H$_2$CO, CH$_2$OH, CH$_3$CHO, CH$_3$OCH$_3$, CH$_3$CH$_2$OH, OHCH$_2$CH$_2$OH, and
a mixture of complex CHO- and COOH-containing molecules (HCOOH, HCOOCH$_3$, OHCH$_2$CHO)
after UV photolysis of CH$_3$OH-rich ices.
Nonetheless, they found 
a comparatively small formation of CH$_3$O-containing molecules during
warm-up of the irradiated ice.

Owing to the large CH$_3$OH abundance found
in the present work for all the studied components, we expect CH$_3$OH-rich ices
in our source. However, we note that other processes different from UV radiation
which deposit energy to the dust grain (i.e. cosmic ray-grain interactions,
grain-grain collisions, IR heating) could lead to different products.

\item Chemical models provided by \citet{Garrod_2008} show that a wide array of complex species may
be formed by reactions involving radicals on dust grains. The reaction
of the mobile primary radicals HCO and CH$_3$ with more
strongly bound primary radicals CH$_3$O and CH$_2$OH results in
the formation of HCOOCH$_3$, CH$_3$OCH$_3$, OHCH$_2$CHO,
and CH$_3$CH$_2$OH (i.e. both C$-$O$-$C- and C$-$O$-$H-containing molecules)
at temperatures between 30 and 40\,K. For the most complex C$-$O$-$C-bearing molecules, 
\citet{Belloche_2009} proposed that HCOOCH$_2$CH$_3$ is 
primarily formed on grains by adding HCO or CH$_3$ to functional-group
radicals derived from HCOOCH$_3$ and CH$_3$CH$_2$OH. 

\end{itemize}
\vspace{-\topsep}
With regard to gas chemistry, several gas-phase mechanisms have been proposed to
explain the high abundances of several C$-$O$-$C-bearing molecules:
\vspace{-\topsep}
\begin{itemize}
\item 
To produce HCOOCH$_3$, \citet{Neill_2012} proposed ion-molecule reactions
in the gas phase that involve the reaction of CH$_3$OH and HCOOH.
\item 
Furthermore, \citet{Balucani_2015} proposed the non-thermal desorption of iced methanol
in cold environments as the precursor of a series of
gas-phase reactions, some of them being uncertain radiative association reactions,
leading to the formation of CH$_3$O, CH$_3$OCH$_3$, and
HCOOCH$_3$. 
\item \citet{Taquet_2016} explained the abundances of CH$_3$OCH$_3$,
HCOOCH$_3$, HCOOCH$_2$CH$_3$, and CH$_3$CH$_2$OCH$_3$ found
in several star-forming regions by means of ion-neutral gas-phase chemistry triggered by the evaporation of interstellar
ices at temperatures higher than 100\,K. The gas-phase chemical network used by these authors
assumes CH$_3$OH as the main precursor to form CH$_3$OCH$_3$ and HCOOCH$_3$,
and CH$_3$CH$_2$OH as the starting point to produce the most complex species (HCOOCH$_2$CH$_3$ and CH$_3$CH$_2$OCH$_3$).
\end{itemize}
\vspace{-\topsep}
Thus, we could expect that reactions in the gas phase have led to decrease the abundances of C$-$O$-$H-containing
molecules (which emerge in the gas phase more likely by desorption from the ices)
and enhance the abundance of C$-$O$-$C-containing species. If we assume
that the released species from the grain mantles are 
similar for both the compact ridge and the south hot core,
the decreased abundance of CH$_3$CH$_2$OH in the former could reflect the formation 
of HCOOCH$_2$CH$_3$ and CH$_3$CH$_2$OCH$_3$ via the mechanisms proposed by \citet{Taquet_2016}.
This could point out to an advanced evolutionary stage for the compact ridge component.

As an example of the constraints brought by the spatial distribution of the species,
it seems that of the two paths to form HCOOCH$_2$CH$_3$ proposed by \citet{Belloche_2009}, the one implying
a role of HCOOCH$_3$ is favoured in Orion KL. Namely the distribution of HCOOCH$_2$CH$_3$ resembles that of
HCOOCH$_3$, and not that of CH$_3$CH$_2$OH. Furthermore, HCOOCH$_3$ could also be similarly the precursor of
CH$_3$COOCH$_3$ on the grain surface through
the COOCH$_3$ radical.

\subsection{Warm-up timescales and kinetic temperatures}

On a smaller scale, the two different emission peaks found in the hot core south (EG and ET peaks) 
may reflect different warm-up timescales and/or different kinetic temperatures of the gas.

\citet{Garrod_2008} also proposed that 
the formation of CH$_3$OCH$_2$OH and OHCH$_2$CH$_2$OH 
is dependent on the mobility of the CH$_3$O and CH$_2$OH radicals, respectively.
CH$_2$OH becomes mobile just as water and other species are beginning to desorb. 
The grain surface OHCH$_2$CH$_2$OH abundance rises dramatically at 110 K.
At temperatures between 50 and 100\,K, secondary radicals
themselves may become mobile; high temperature mobility of
the secondary radical CH$_3$CO results in reactions with CH$_3$ and OH to form
CH$_3$COCH$_3$ and CH$_3$COOH, respectively. 
The formation of these secondary radical products is
most effective with long warm-up timescales. 
Therefore,  in the ET peak CH$_3$CH$_2$OH and CH$_3$OCH$_2$OH are desorbed after reaction of primary radicals; instead, in the EG peak the low mobility of the CH$_2$OH radical could lead to the desorption of OHCH$_2$CH$_2$OH
at similar physical conditions of the CH$_3$COOH desorption which is produced after 
reactions involving secondary radicals. 

Interestingly, to explain the high abundance found for 
CH$_3$OCH$_2$OH in NGC\,6334I, \citet{McGuire2017} suggest an important role of low-energy electrons
(as a product of the interactions of high-energy radiation, such as cosmic
rays, with matter) acting on methanol ices, as studied by \citet{Sullivan_2016}.

\subsection{Desorption mechanisms}
Different desorption mechanisms could favour the enhancement of different species in the gas phase.
The core of the EG peak could be more likely affected by desorption mechanisms via non-thermal processes due to its
embedded position in the SiO outflow. 
In this scenario, sputtering and grain collisions could favour
the production of species which, in the mechanism proposed by \citet{Garrod_2008},
are produced when the dust grain reaches high temperatures and certain radicals (i.e. CH$_2$OH and
secondary radicals) become mobile. Then,
OHCH$_2$CH$_2$OH and CH$_3$COOH could be enhanced into the gas phase   
before the species that warm up at lower temperatures (i.e. HCOOCH$_3$, CH$_3$OCH$_3$, OHCH$_2$CHO, CH$_3$CH$_2$OH)
reach high abundances. This also explains the spatial distribution of CH$_3$COCH$_3$ as produced by
the interaction of a front shock with its surroundings. 
The lower diffusion barrier of the CH$_3$ radical (compared to the OH radical)
may also explain the differences found in the spatial distribution of CH$_3$COOH and CH$_3$COCH$_3$.

\subsection{Early stage of shock chemistry}
The co-spatial emission of HCOOH, CH$_3$COOH, and OHCH$_2$CH$_2$OH may indicate
an early stage of shock chemistry.
HCOOH may be easily liberated from grain surfaces \citep{Garrod_2008}, but it could
be processed into HCOOCH$_3$ in regions where the timescales are sufficiently 
long for gas-phase reactions to occur \citep{Neill_2012}. This also explains the decreased emission of
HCOOH in the MF and ET peaks. In this scenario, the chemistry found in MF and ET peaks
could reflect a prevailing role of gas-phase reactions that take place after sputtering (in a post-shock scenario)
and/or thermal evaporation (by compression of the gas in a pre-shock scenario) of the grain mantles.
In this respect, if the first generation of mantle molecules might include $-$OH-bearing species, the first group of species (from B to G)
will be characterised by radicals where the OH bond has been cut (not only CH$_3$O, but CH$_3$CO to form CH$_3$COOCH$_3$, 
and CH$_3$CH$_2$O to form HCOOCH$_2$CH$_3$ and CH$_3$CH$_2$OCH$_3$).
\vspace{\topsep}

To sum up, we found an evident difference between the molecular content of the compact ridge and the hot core south,
which is reflected in the internal chemical structure of the detected species.
This observed spatial differentiation between the most complex O-bearing species could provide an important constraint to future investigations regarding 
the chemistry of the interstellar medium. Chemical models will have to be adapted to these new
results.
Finally, to explain this chemical segregation issues such as the evolutionary stage of the different Orion~KL components, 
the composition of the dust grains prior to the collapse phase, the
molecules accreted onto the grains, desorption mechanisms in the different regions (thermal/chemical/sputtering/UV processes),
and/or the surface and gas-phase chemistry should be addressed in a combined way.


\begin{acknowledgements} 
We would like to thank the anonymous referee for a helpful
report that led to improvements in the paper.
The authors thank R.\,Motiyenko, L.\,Margul\`es, and J.-C.\,Guillemin for their work on methoxymethanol,
which was absolutely key to its detection in Orion KL.
We thank the ERC for support under grant ERC-2013-Syg-610256- NANOCOSMOS. We also thank
the Spanish MINECO for funding support under grants AYA2012-32032 and FIS2014-52172-C2,
and  the CONSOLIDER-Ingenio programme $``$ASTROMOL$"$ CSD 2009-00038.
\end{acknowledgements}


\bibliographystyle{aa}
\bibliography{references}

\begin{thebibliography}{53}
\expandafter\ifx\csname natexlab\endcsname\relax\def\natexlab#1{#1}\fi

\bibitem[{{Allen} {et~al.}(2017){Allen}, {van der Tak}, {S{\'a}nchez-Monge},
  {Cesaroni}, \& {Beltr{\'a}n}}]{Allen_2017}
{Allen}, V., {van der Tak}, F.~F.~S., {S{\'a}nchez-Monge}, {\'A}., {Cesaroni},
  R., \& {Beltr{\'a}n}, M.~T. 2017, \aap, 603, A133

\bibitem[{{Bally} {et~al.}(2017){Bally}, {Ginsburg}, {Arce}, {Eisner},
  {Youngblood}, {Zapata}, \& {Zinnecker}}]{Bally_2017}
{Bally}, J., {Ginsburg}, A., {Arce}, H., {et~al.} 2017, \apj, 837, 60

\bibitem[{{Balucani} {et~al.}(2015){Balucani}, {Ceccarelli}, \&
  {Taquet}}]{Balucani_2015}
{Balucani}, N., {Ceccarelli}, C., \& {Taquet}, V. 2015, \mnras, 449, L16

\bibitem[{{Bell} {et~al.}(2014){Bell}, {Cernicharo}, {Viti}, {Marcelino},
  {Palau}, {Esplugues}, \& {Tercero}}]{Bell_2014}
{Bell}, T.~A., {Cernicharo}, J., {Viti}, S., {et~al.} 2014, \aap, 564, A114

\bibitem[{{Belloche} {et~al.}(2009){Belloche}, {Garrod}, {M{\"u}ller},
  {Menten}, {Comito}, \& {Schilke}}]{Belloche_2009}
{Belloche}, A., {Garrod}, R.~T., {M{\"u}ller}, H.~S.~P., {et~al.} 2009, \aap,
  499, 215

\bibitem[{{Bergner} {et~al.}(2017){Bergner}, {{\"O}berg}, {Garrod}, \&
  {Graninger}}]{Bergner_2017}
{Bergner}, J.~B., {{\"O}berg}, K.~I., {Garrod}, R.~T., \& {Graninger}, D.~M.
  2017, \apj, 841, 120

\bibitem[{{Blake} {et~al.}(1987){Blake}, {Sutton}, {Masson}, \&
  {Phillips}}]{Blake_1987}
{Blake}, G.~A., {Sutton}, E.~C., {Masson}, C.~R., \& {Phillips}, T.~G. 1987,
  \apj, 315, 621

\bibitem[{{Brouillet} {et~al.}(2013){Brouillet}, {Despois}, {Baudry}, {Peng},
  {Favre}, {Wootten}, {Remijan}, {Wilson}, {Combes}, \&
  {Wlodarczak}}]{Brouillet_2013}
{Brouillet}, N., {Despois}, D., {Baudry}, A., {et~al.} 2013, \aap, 550, A46

\bibitem[{{Brouillet} {et~al.}(2015){Brouillet}, {Despois}, {Lu}, {Baudry},
  {Cernicharo}, {Bockel{\'e}e-Morvan}, {Crovisier}, \&
  {Biver}}]{Brouillet_2015}
{Brouillet}, N., {Despois}, D., {Lu}, X.-H., {et~al.} 2015, \aap, 576, A129

\bibitem[{{Cernicharo}(2012)}]{Cernicharo_2012}
{Cernicharo}, J. 2012, in EAS Publications Series, Vol.~58, EAS Publications
  Series, ed. C.~{Stehl{\'e}}, C.~{Joblin}, \& L.~{d'Hendecourt}, 251--261

\bibitem[{{Cernicharo} {et~al.}(2016){Cernicharo}, {Kisiel}, {Tercero},
  {Kolesnikov{\'a}}, {Medvedev}, {L{\'o}pez}, {Fortman}, {Winnewisser}, {de
  Lucia}, {Alonso}, \& {Guillemin}}]{Cernicharo_2016}
{Cernicharo}, J., {Kisiel}, Z., {Tercero}, B., {et~al.} 2016, \aap, 587, L4

\bibitem[{{Crockett} {et~al.}(2014){Crockett}, {Bergin}, {Neill}, {Favre},
  {Schilke}, {Lis}, {Bell}, {Blake}, {Cernicharo}, {Emprechtinger},
  {Esplugues}, {Gupta}, {Kleshcheva}, {Lord}, {Marcelino}, {McGuire},
  {Pearson}, {Phillips}, {Plume}, {van der Tak}, {Tercero}, \&
  {Yu}}]{Crockett_2014}
{Crockett}, N.~R., {Bergin}, E.~A., {Neill}, J.~L., {et~al.} 2014, \apj, 787,
  112

\bibitem[{{Favre} {et~al.}(2011{\natexlab{a}}){Favre}, {Despois}, {Brouillet},
  {Baudry}, {Combes}, {Gu{\'e}lin}, {Wootten}, \& {Wlodarczak}}]{Favre_2011a}
{Favre}, C., {Despois}, D., {Brouillet}, N., {et~al.} 2011{\natexlab{a}}, \aap,
  532, A32

\bibitem[{{Favre} {et~al.}(2017){Favre}, {Pagani}, {Goldsmith}, {Bergin},
  {Carvajal}, {Kleiner}, {Melnick}, \& {Snell}}]{Favre_2017}
{Favre}, C., {Pagani}, L., {Goldsmith}, P.~F., {et~al.} 2017, \aap, 604, L2

\bibitem[{{Favre} {et~al.}(2011{\natexlab{b}}){Favre}, {Wootten}, {Remijan},
  {Brouillet}, {Wilson}, {Despois}, \& {Baudry}}]{Favre_2011b}
{Favre}, C., {Wootten}, H.~A., {Remijan}, A.~J., {et~al.} 2011{\natexlab{b}},
  \apjl, 739, L12

\bibitem[{{Fayolle} {et~al.}(2015){Fayolle}, {{\"O}berg}, {Garrod}, {van
  Dishoeck}, \& {Bisschop}}]{Fayolle_2015}
{Fayolle}, E.~C., {{\"O}berg}, K.~I., {Garrod}, R.~T., {van Dishoeck}, E.~F.,
  \& {Bisschop}, S.~E. 2015, \aap, 576, A45

\bibitem[{{Feng} {et~al.}(2015){Feng}, {Beuther}, {Henning}, {Semenov},
  {Palau}, \& {Mills}}]{Feng_2015}
{Feng}, S., {Beuther}, H., {Henning}, T., {et~al.} 2015, \aap, 581, A71

\bibitem[{{Friedel} \& {Widicus Weaver}(2012)}]{Friedel_2012}
{Friedel}, D.~N. \& {Widicus Weaver}, S.~L. 2012, \apjs, 201, 17

\bibitem[{{Garrod} {et~al.}(2008){Garrod}, {Widicus Weaver}, \&
  {Herbst}}]{Garrod_2008}
{Garrod}, R.~T., {Widicus Weaver}, S.~L., \& {Herbst}, E. 2008, \apj, 682, 283

\bibitem[{{Genzel} \& {Stutzki}(1989)}]{Genzel_1989}
{Genzel}, R. \& {Stutzki}, J. 1989, \araa, 27, 41

\bibitem[{{Goicoechea} {et~al.}(2015){Goicoechea}, {Chavarr{\'{\i}}a},
  {Cernicharo}, {Neufeld}, {Vavrek}, {Bergin}, {Cuadrado}, {Encrenaz},
  {Etxaluze}, {Melnick}, \& {Polehampton}}]{Goicoechea_2015}
{Goicoechea}, J.~R., {Chavarr{\'{\i}}a}, L., {Cernicharo}, J., {et~al.} 2015,
  \apj, 799, 102

\bibitem[{{G{\'o}mez} {et~al.}(2005){G{\'o}mez}, {Rodr{\'{\i}}guez}, {Loinard},
  {Lizano}, {Poveda}, \& {Allen}}]{Gomez_2005}
{G{\'o}mez}, L., {Rodr{\'{\i}}guez}, L.~F., {Loinard}, L., {et~al.} 2005, \apj,
  635, 1166

\bibitem[{{Grossschedl} {et~al.}(2018){Grossschedl}, {Alves}, {Meingast},
  {Ackerl}, {Ascenso}, {Bouy}, {Burkert}, {Forbrich}, {Fuernkranz}, {Goodman},
  {Hacar}, {Herbst-Kiss}, {Lada}, {Larreina}, {Leschinski}, {Lombardi},
  {Moitinho}, {Mortimer}, \& {Zari}}]{Grossschedl_2018}
{Grossschedl}, J.~E., {Alves}, J., {Meingast}, S., {et~al.} 2018, ArXiv
  e-prints

\bibitem[{{Herbst} \& {van Dishoeck}(2009)}]{Herbst_2009}
{Herbst}, E. \& {van Dishoeck}, E.~F. 2009, \araa, 47, 427

\bibitem[{{Hirota} {et~al.}(2015){Hirota}, {Kim}, {Kurono}, \&
  {Honma}}]{Hirota_2015}
{Hirota}, T., {Kim}, M.~K., {Kurono}, Y., \& {Honma}, M. 2015, \apj, 801, 82

\bibitem[{{H{\"o}gbom}(1974)}]{Hogbom_1974}
{H{\"o}gbom}, J.~A. 1974, \aaps, 15, 417

\bibitem[{{Jim{\'e}nez-Serra} {et~al.}(2016){Jim{\'e}nez-Serra}, {Vasyunin},
  {Caselli}, {Marcelino}, {Billot}, {Viti}, {Testi}, {Vastel}, {Lefloch}, \&
  {Bachiller}}]{JimenezSerra_2016}
{Jim{\'e}nez-Serra}, I., {Vasyunin}, A.~I., {Caselli}, P., {et~al.} 2016,
  \apjl, 830, L6

\bibitem[{{Kounkel} {et~al.}(2017){Kounkel}, {Hartmann}, {Loinard},
  {Ortiz-Le{\'o}n}, {Mioduszewski}, {Rodr{\'{\i}}guez}, {Dzib}, {Torres},
  {Pech}, {Galli}, {Rivera}, {Boden}, {Evans}, {Brice{\~n}o}, \&
  {Tobin}}]{Kounkel_2017}
{Kounkel}, M., {Hartmann}, L., {Loinard}, L., {et~al.} 2017, \apj, 834, 142

\bibitem[{{Luhman} {et~al.}(2017){Luhman}, {Robberto}, {Tan}, {Andersen},
  {Giulia Ubeira Gabellini}, {Manara}, {Platais}, \& {Ubeda}}]{Luhman_2017}
{Luhman}, K.~L., {Robberto}, M., {Tan}, J.~C., {et~al.} 2017, \apjl, 838, L3

\bibitem[{{McGuire} {et~al.}(2017){McGuire}, {Shingledecker}, {Willis},
  {Burkhardt}, {El-Abd}, {Motiyenko}, {Brogan}, {Hunter}, {Margul{\`e}s},
  {Guillemin}, {Garrod}, {Herbst}, \& {Remijan}}]{McGuire2017}
{McGuire}, B.~A., {Shingledecker}, C.~N., {Willis}, E.~R., {et~al.} 2017,
  \apjl, 851, L46

\bibitem[{{Menten} {et~al.}(2007){Menten}, {Reid}, {Forbrich}, \&
  {Brunthaler}}]{Menten_2007}
{Menten}, K.~M., {Reid}, M.~J., {Forbrich}, J., \& {Brunthaler}, A. 2007, \aap,
  474, 515

\bibitem[{{Motiyenko} {et~al.}(2018){Motiyenko}, {Margul{\`e}s}, {Despois}, \&
  {Guillemin}}]{Motiyenko_2018}
{Motiyenko}, R.~A., {Margul{\`e}s}, L., {Despois}, D., \& {Guillemin}, J.-C.
  2018, Physical Chemistry Chemical Physics (Incorporating Faraday
  Transactions), 20, 5509

\bibitem[{{Neill} {et~al.}(2012){Neill}, {Muckle}, {Zaleski}, {Steber}, {Pate},
  {Lattanzi}, {Spezzano}, {McCarthy}, \& {Remijan}}]{Neill_2012}
{Neill}, J.~L., {Muckle}, M.~T., {Zaleski}, D.~P., {et~al.} 2012, \apj, 755,
  153

\bibitem[{{Neill} {et~al.}(2013){Neill}, {Wang}, {Bergin}, {Crockett}, {Favre},
  {Plume}, \& {Melnick}}]{Neill_2013}
{Neill}, J.~L., {Wang}, S., {Bergin}, E.~A., {et~al.} 2013, \apj, 770, 142

\bibitem[{{{\"O}berg} {et~al.}(2013){{\"O}berg}, {Boamah}, {Fayolle}, {Garrod},
  {Cyganowski}, \& {van der Tak}}]{Oberg_2013}
{{\"O}berg}, K.~I., {Boamah}, M.~D., {Fayolle}, E.~C., {et~al.} 2013, \apj,
  771, 95

\bibitem[{{{\"O}berg} {et~al.}(2009){{\"O}berg}, {Garrod}, {van Dishoeck}, \&
  {Linnartz}}]{Oberg_2009}
{{\"O}berg}, K.~I., {Garrod}, R.~T., {van Dishoeck}, E.~F., \& {Linnartz}, H.
  2009, \aap, 504, 891

\bibitem[{{O'Dell}(2001)}]{ODell_2001}
{O'Dell}, C.~R. 2001, \pasp, 113, 29

\bibitem[{{Pagani} {et~al.}(2017){Pagani}, {Favre}, {Goldsmith}, {Bergin},
  {Snell}, \& {Melnick}}]{Pagani_2017}
{Pagani}, L., {Favre}, C., {Goldsmith}, P.~F., {et~al.} 2017, \aap, 604, A32

\bibitem[{{Peng} {et~al.}(2013){Peng}, {Despois}, {Brouillet}, {Baudry},
  {Favre}, {Remijan}, {Wootten}, {Wilson}, {Combes}, \&
  {Wlodarczak}}]{Peng_2013}
{Peng}, T.-C., {Despois}, D., {Brouillet}, N., {et~al.} 2013, \aap, 554, A78

\bibitem[{{Peng} {et~al.}(2012){Peng}, {Despois}, {Brouillet}, {Parise}, \&
  {Baudry}}]{Peng_2012}
{Peng}, T.-C., {Despois}, D., {Brouillet}, N., {Parise}, B., \& {Baudry}, A.
  2012, \aap, 543, A152

\bibitem[{{Rodr{\'{\i}}guez} {et~al.}(2017){Rodr{\'{\i}}guez}, {Dzib},
  {Loinard}, {Zapata}, {G{\'o}mez}, {Menten}, \& {Lizano}}]{Rodriguez_2017}
{Rodr{\'{\i}}guez}, L.~F., {Dzib}, S.~A., {Loinard}, L., {et~al.} 2017, \apj,
  834, 140

\bibitem[{{Schilke} {et~al.}(2001){Schilke}, {Benford}, {Hunter}, {Lis}, \&
  {Phillips}}]{Schilke_2001}
{Schilke}, P., {Benford}, D.~J., {Hunter}, T.~R., {Lis}, D.~C., \& {Phillips},
  T.~G. 2001, \apjs, 132, 281

\bibitem[{{Sullivan} {et~al.}(2016){Sullivan}, {Boamah}, {Shulenberger},
  {Chapman}, {Atkinson}, {Boyer}, \& {Arumainayagam}}]{Sullivan_2016}
{Sullivan}, K.~K., {Boamah}, M.~D., {Shulenberger}, K.~E., {et~al.} 2016,
  \mnras, 460, 664

\bibitem[{{Suzuki} {et~al.}(2018){Suzuki}, {Ohishi}, {Saito}, {Hirota},
  {Majumdar}, \& {Wakelam}}]{Suzuki_2018}
{Suzuki}, T., {Ohishi}, M., {Saito}, M., {et~al.} 2018, \apjs, 237, 3

\bibitem[{{Taquet} {et~al.}(2016){Taquet}, {Wirstr{\"o}m}, \&
  {Charnley}}]{Taquet_2016}
{Taquet}, V., {Wirstr{\"o}m}, E.~S., \& {Charnley}, S.~B. 2016, \apj, 821, 46

\bibitem[{{Tercero} {et~al.}(2015){Tercero}, {Cernicharo}, {L{\'o}pez},
  {Brouillet}, {Kolesnikov{\'a}}, {Motiyenko}, {Margul{\`e}s}, {Alonso}, \&
  {Guillemin}}]{Tercero_2015}
{Tercero}, B., {Cernicharo}, J., {L{\'o}pez}, A., {et~al.} 2015, \aap, 582, L1

\bibitem[{{Tercero} {et~al.}(2010){Tercero}, {Cernicharo}, {Pardo}, \&
  {Goicoechea}}]{Tercero_2010}
{Tercero}, B., {Cernicharo}, J., {Pardo}, J.~R., \& {Goicoechea}, J.~R. 2010,
  \aap, 517, A96

\bibitem[{{Tercero} {et~al.}(2011){Tercero}, {Vincent}, {Cernicharo}, {Viti},
  \& {Marcelino}}]{Tercero_2011}
{Tercero}, B., {Vincent}, L., {Cernicharo}, J., {Viti}, S., \& {Marcelino}, N.
  2011, \aap, 528, A26

\bibitem[{{Widicus Weaver} \& {Friedel}(2012)}]{Widicus_2012}
{Widicus Weaver}, S.~L. \& {Friedel}, D.~N. 2012, \apjs, 201, 16

\bibitem[{{Wright} \& {Plambeck}(2017)}]{Wright_2017}
{Wright}, M.~C.~H. \& {Plambeck}, R.~L. 2017, \apj, 843, 83

\bibitem[{{Wu} {et~al.}(2014){Wu}, {Liu}, \& {Qin}}]{Wu_2014}
{Wu}, Y., {Liu}, T., \& {Qin}, S.-L. 2014, \apj, 791, 123

\bibitem[{{Zapata} {et~al.}(2011){Zapata}, {Loinard}, {Schmid-Burgk},
  {Rodr{\'{\i}}guez}, {Ho}, \& {Patel}}]{Zapata_2011}
{Zapata}, L.~A., {Loinard}, L., {Schmid-Burgk}, J., {et~al.} 2011, \apjl, 726,
  L12

\bibitem[{{Zapata} {et~al.}(2009){Zapata}, {Schmid-Burgk}, {Ho},
  {Rodr{\'{\i}}guez}, \& {Menten}}]{Zapata_2009}
{Zapata}, L.~A., {Schmid-Burgk}, J., {Ho}, P.~T.~P., {Rodr{\'{\i}}guez}, L.~F.,
  \& {Menten}, K.~M. 2009, \apjl, 704, L45

\end{thebibliography}


\begin{appendix}

\section{Table of rotational transitions}

\begin{table*}
\caption{List of the selected rotational transitions of each O-bearing COM plotted in Fig.~\ref{fig_maps}.}
\label{table_lines}     
\begin{tabular}{c l l c r c l c@{\vrule height 10pt depth 5pt width 0pt}} 
\hline\hline                                                                                  

    Map &      Molecule           & \multicolumn{1}{c}{Transition}                & Frequency    & \multicolumn{1}{c}{$E_{\rm upp}$}  & $A_{\rm ul}$     & \multicolumn{1}{c}{Scale} & Notes    \\
        &                         &                           &  [MHz]       &  \multicolumn{1}{c}{[K]}           & [s$^{-1}$]                  &       
        \multicolumn{1}{c}{[Jy km beam$^{-1}$ s$^{-1}$]}  &      \\
\hline                                                                                                                                          
A       &   $^{13}$CH$_3$OH       &   8$_{4,4}$\,$\rightarrow$\,9$_{3,7}$; A   &  215722.475  &   162.3        &   9.1\,$\times$\,10$^{-6}$  & colour: 0.05 to 3.5   & \\ \cline{3-8} 
        &                         &   5$_{-2,4}$\,$\rightarrow$\,4$_{-2,3}$; E &  236062.000  &    52.1        &   3.8\,$\times$\,10$^{-5}$  & contour: 0.05 to 36 by 3.0  & \\   
        &                         &   5$_{2,3}$\,$\rightarrow$\,4$_{2,2}$; E   &  236062.850  &    48.5        &   3.8\,$\times$\,10$^{-5}$  &              &           \\   
\hline                                                                                                                                         
B       &     HCOOCH$_3$          &   10$_{ 4,6}$\,$\rightarrow$\,9$_{3,7}$; A      &  221717.125  &   43.2        &   1.0\,$\times$\,10$^{-5}$  & colour: 0.05 to 6.5  &   \\ \cline{3-8}
        &                         &   27$_{-9,18}$\,$\rightarrow$\,27$_{-8,19}$; E  &  222367.071  &  277.3        &   1.5\,$\times$\,10$^{-5}$  & contour: 0.05 to 3.6 by 0.3 &   \\   
\hline                                                                                                                                        
C       &   CH$_3$OCH$_3$         &   14$_{1,13}$\,$\rightarrow$\,13$_{2,12}$; AA   &  226346.051    &  98.9     &  3.3\,$\times$\,10$^{-5}$  & colour: 0.07 to 25  &  a  \\
        &                         &   14$_{1,13}$\,$\rightarrow$\,13$_{2,12}$; EE   &  226346.927    &  98.9     &  3.3\,$\times$\,10$^{-5}$  &             &    \\
        &                         &   14$_{1,13}$\,$\rightarrow$\,13$_{2,12}$; AE   &  226347.803    &  98.9     &  3.3\,$\times$\,10$^{-5}$  &             &    \\
        &                         &   14$_{1,13}$\,$\rightarrow$\,13$_{2,12}$; EA   &  226347.804    &  98.9     &  3.3\,$\times$\,10$^{-5}$  &             &    \\   \cline{3-8}
        &                         &   24$_{4,21}$\,$\rightarrow$\,24$_{3,22}$; EA   &  225202.460    &  296.4     &  4.7\,$\times$\,10$^{-5}$  & contour: 0.07 to 15 by 1.0  &   \\
        &                         &   24$_{4,21}$\,$\rightarrow$\,24$_{3,22}$; AE   &  225202.460    &  296.4     &  4.7\,$\times$\,10$^{-5}$  &            &              \\
        &                         &   24$_{4,21}$\,$\rightarrow$\,24$_{3,22}$; EE   &  225203.946    &  296.4     &  4.7\,$\times$\,10$^{-5}$  &            &              \\
        &                         &   24$_{4,21}$\,$\rightarrow$\,24$_{3,22}$; AA   &  225205.431    &  296.4     &  4.7\,$\times$\,10$^{-5}$  &            &              \\   
\hline                                                                                                                                                    
D       &   CH$_2$OCH$_2$         &   7$_{2,6}$\,$\rightarrow$\,6$_{1,5}$; ortho    &  226072.102  &    47.0      &   1.7\,$\times$\,10$^{-4}$  & colour: 0.10 to 5.5  & b \\    \cline{3-8}
        &                         &  18$_{6,12}$\,$\rightarrow$\,18$_{5,13}$; ortho &  234979.289  &   321.6      &   9.2\,$\times$\,10$^{-5}$  & contour: 0.05 to 1.7 by 0.15 &  \\ 
        &                         &  18$_{7,12}$\,$\rightarrow$\,18$_{6,13}$; para  &  234979.462  &   319.9      &   9.2\,$\times$\,10$^{-5}$  &            &            \\ 
\hline                                                                                                                                                        
E       &   CH$_3$COOCH$_3$       &  18$_{-9,9}$\,$\rightarrow$\,17$_{-8,9}$; EA  &   242955.602 &    86.3        &   1.0\,$\times$\,10$^{-4}$  & colour: 0.05 to 0.65  & c  \\ \cline{3-8}
        &                         &   14$_{9,6}$\,$\rightarrow$\,13$_{8,6}$; AE   &   216462.024 &    63.0        &   9.2\,$\times$\,10$^{-5}$  & contour: 0.05 to 0.35 by 0.05&   \\  
\hline                                                                                                                                          
F       &      HCOOCH$_2$CH$_3$   &  43$_{6,38}$\,$\rightarrow$\,42$_{6,37}$; trans   &   237566.732 &   275.9       &  2.6\,$\times$\,10$^{-4}$   & colour: 0.05 to 0.40    &  d   \\  \cline{3-8}
        &                         &  43$_{1,42}$\,$\rightarrow$\,42$_{1,41}$; trans   &   228399.191 &   247.0       &  2.3\,$\times$\,10$^{-4}$   & contour: 0.05 to 0.40 by 0.05  &     e \\ 
\hline                                                                                                                                           
G       &   CH$_3$CH$_2$OCH$_3$   &  31$_{1,31}$\,$\rightarrow$\,30$_{0,30}$; EE'  &   245274.088 &   188.8       &  9.5\,$\times$\,10$^{-5}$   & colour: 0.02 to 0.90   &  f \\                                                             
        &                         &  31$_{1,31}$\,$\rightarrow$\,30$_{0,30}$; EE   &   245274.088 &   188.8       &  9.5\,$\times$\,10$^{-5}$   &              &  \\
        &                         &  31$_{1,31}$\,$\rightarrow$\,30$_{0,30}$; AE   &   245274.098 &   188.8       &  4.8\,$\times$\,10$^{-5}$   &              &  \\
        &                         &  31$_{1,31}$\,$\rightarrow$\,30$_{0,30}$; EA   &   245274.211 &   188.8       &  4.8\,$\times$\,10$^{-5}$   &              &  \\
        &                         &  31$_{1,31}$\,$\rightarrow$\,30$_{0,30}$; AA   &   245274.221 &   188.8       &  4.8\,$\times$\,10$^{-5}$   &              &  \\   \cline{3-8}
        &                         &  23$_{2,22}$\,$\rightarrow$\,22$_{1,21}$; EE'  &   225494.508 &   111.0       &  3.4\,$\times$\,10$^{-5}$   & contour: 0.05 to 0.35 by 0.05  & g \\ 
        &                         &  23$_{2,22}$\,$\rightarrow$\,22$_{1,21}$; EE   &   225494.508 &   111.0       &  3.4\,$\times$\,10$^{-5}$   &              &           \\ 
        &                         &  23$_{2,22}$\,$\rightarrow$\,22$_{1,21}$; AE   &   225494.625 &   111.0       &  1.7\,$\times$\,10$^{-5}$   &              &           \\ 
        &                         &  23$_{2,22}$\,$\rightarrow$\,22$_{1,21}$; EA   &   225495.848 &   111.0       &  1.7\,$\times$\,10$^{-5}$   &              &           \\ 
        &                         &  23$_{2,22}$\,$\rightarrow$\,22$_{1,21}$; AA   &   225495.965 &   111.0       &  1.7\,$\times$\,10$^{-5}$   &              &           \\ 
\hline                                                                                                                                     
H       &   HCOOH                 &  10$_{7, 4}$\,$\rightarrow$\,9$_{7,3}$     &   224911.799 &   215.0       &  6.5\,$\times$\,10$^{-5}$   & colour: 0.05 to 1.7   &    \\
        &                         &  10$_{7, 3}$\,$\rightarrow$\,9$_{7,2}$     &   224911.799 &   215.0       &  6.5\,$\times$\,10$^{-5}$   &              &              \\
        &                         &  10$_{8, 2}$\,$\rightarrow$\,9$_{8,1}$     &   224911.894 &   262.6       &  4.6\,$\times$\,10$^{-5}$   &              &              \\
        &                         &  10$_{8, 3}$\,$\rightarrow$\,9$_{8,2}$     &   224911.894 &   262.6       &  4.6\,$\times$\,10$^{-5}$   &              &              \\   \cline{3-8}
        &                         &  10$_{0,10}$\,$\rightarrow$\,9$_{0,9}$     &   220037.942 &    58.6       &  1.2\,$\times$\,10$^{-4}$   & contour: 0.05 to 3.2 by 0.20  &              \\  
\hline                                                                                                                                                  
I       &      OHCH$_2$CH$_2$OH   &  23$_{9,15}$\,$\rightarrow$\,22$_{9,14}$; ag   &   242947.990 &   176.1       &  3.0\,$\times$\,10$^{-4}$   & colour: 0.05 to 2.3  &  h  \\                                                
        &                         &  23$_{9,14}$\,$\rightarrow$\,22$_{9,13}$; ag   &   242948.591 &   176.1       &  3.0\,$\times$\,10$^{-4}$   &           &    \\    \cline{3-8}
        &                         &  23$_{2,22}$\,$\rightarrow$\,22$_{2,21}$; ag   &   228752.776 &   132.8       &  2.9\,$\times$\,10$^{-4}$   & contour: 0.05 to 2.0 by 0.20 &  i  \\ 
\hline                                                                                                                                                     
J       &     CH$_3$COOH          &  20$_{0,20}$\,$\rightarrow$\,19$_{0,19}$; E   &   218010.014 &   112.1       &  1.7\,$\times$\,10$^{-4}$   & colour: 0.05 to 1.8  &  j          \\                                               
        &                         &  20$_{0,20}$\,$\rightarrow$\,19$_{1,19}$; E   &   218010.014 &   112.1       &  2.3\,$\times$\,10$^{-8}$   &             &            \\                                                                               
        &                         &  20$_{1,20}$\,$\rightarrow$\,19$_{0,19}$; E   &   218010.014 &   118.8       &  2.3\,$\times$\,10$^{-8}$   &             &            \\
        &                         &  20$_{1,20}$\,$\rightarrow$\,19$_{1,19}$; E   &   218010.014 &   118.8       &  1.7\,$\times$\,10$^{-4}$   &             &            \\    \cline{3-8}                                                                             
        &                         &  22$_{0,22}$\,$\rightarrow$\,21$_{1,21}$; E   &   239305.851 &   134.6       &  4.4\,$\times$\,10$^{-5}$   & contour: 0.05 to 2.5 by 0.20 &            \\                                                                                  
        &                         &  22$_{1,22}$\,$\rightarrow$\,21$_{0,21}$; E   &   239305.851 &   134.6       &  4.4\,$\times$\,10$^{-5}$   &             &            \\                                                                                  
        &                         &  22$_{0,22}$\,$\rightarrow$\,21$_{0,21}$; E   &   239305.851 &   134.6       &  1.8\,$\times$\,10$^{-4}$   &             &            \\                                                                                  
        &                         &  22$_{1,22}$\,$\rightarrow$\,21$_{1,21}$; E   &   239305.851 &   134.6       &  1.8\,$\times$\,10$^{-4}$   &             &            \\                                                                                  
\hline                                                                                                                                                        
\end{tabular}                                                                                                                                   
\end{table*}

\begin{table*}                                                                                                                 
\setcounter{table}{0}                                                                                                                             
\caption{continued.}                                                                                                                                      
\begin{tabular}{c l l c r c l c@{\vrule height 10pt depth 5pt width 0pt}}                                                                         
\hline\hline                                                                                                                                    
                                                                                                                                                
    Map &      Molecule           & \multicolumn{1}{c}{Transition}                & Frequency    & \multicolumn{1}{c}{$E_{\rm upp}$}  & $A_{\rm ul}$       & \multicolumn{1}{c}{Scale}   & Notes   \\
        &                         &                           &  [MHz]       &  \multicolumn{1}{c}{[K]}           & [s$^{-1}$]                  &         
        \multicolumn{1}{c}{[Jy km beam$^{-1}$ s$^{-1}$]}  &      \\  
\hline                                                                                                                                          
K       &   CH$_3$CH$_2$OH        &  20$_{5,15}$\,$\rightarrow$\,20$_{4,16}$; trans   &   222217.285 &   208.3       &  6.5\,$\times$\,10$^{-5}$   & colour: 0.05 to 6.5  &  k\\    \cline{3-8}
        &                         &  8$_{5,3}$\,$\rightarrow$\,8$_{4,4}$; trans       &   234984.238 &    61.6       &  5.8\,$\times$\,10$^{-5}$   & contour: 0.05 to 8.0 by 1.0 &  \\ 
        \hline                                                                                                                                          
L       &   CH$_3$OCH$_2$OH       &  22$_{10,12}$\,$\rightarrow$\,21$_{10,11}$ &   229025.777 &   184.0       &  3.9\,$\times$\,10$^{-6}$   & colour:  0.05 to 2.0   &     \\ 
        &                         &  22$_{10,13}$\,$\rightarrow$\,21$_{10,12}$ &   229027.637 &   184.0       &  3.9\,$\times$\,10$^{-6}$   &                 &      \\ 
        &                         &  22$_{10,12}$\,$\rightarrow$\,21$_{10,11}$ &   229027.637 &   184.0       &  3.9\,$\times$\,10$^{-6}$   &                 &      \\ 
        &                         &  22$_{10,13}$\,$\rightarrow$\,21$_{10,12}$ &   229028.550 &   184.0       &  3.9\,$\times$\,10$^{-6}$   &                 &      \\     \cline{3-8}
        &                         &  23$_{4,19}$\,$\rightarrow$\,22$_{4,18}$   &   246243.055 &   148.9       &  5.9\,$\times$\,10$^{-6}$   & contour: 0.05 to 1.8 by 0.20    &       \\
        &                         &  23$_{4,19}$\,$\rightarrow$\,22$_{4,18}$   &   246246.232 &   148.9       &  5.9\,$\times$\,10$^{-6}$   &                 &       \\
\hline                                                                                                                                         
M       &   OHCH$_2$CHO           &  24$_{0,24}$\,$\rightarrow$\,23$_{1,23}$   &   242957.815 &   148.3       &  4.2\,$\times$\,10$^{-4}$   & colour: 0.05 to 0.70   &  l \\   
        &                         &  24$_{1,24}$\,$\rightarrow$\,23$_{1,23}$   &   242957.904 &   148.3       &  5.6\,$\times$\,10$^{-6}$   &              &  \\ 
        &                         &  24$_{0,24}$\,$\rightarrow$\,23$_{0,23}$   &   242957.983 &   148.3       &  5.6\,$\times$\,10$^{-6}$   &              &  \\ 
        &                         &  24$_{1,24}$\,$\rightarrow$\,23$_{0,23}$   &   242958.072 &   148.3       &  4.2\,$\times$\,10$^{-4}$   &              &  \\   \cline{3-8}
        &                         &  22$_{1,21}$\,$\rightarrow$\,21$_{2,20}$   &   232286.154 &   134.5       &  3.1\,$\times$\,10$^{-4}$   & contour: 0.05 to 0.45 by 0.05  & m  \\
\hline                                                                                                                                         
N       &   CH$_3$COCH$_3$        &  22$_{1,21}$\,$\rightarrow$\,21$_{2,20}$; EE &  229055.797 &   132.8       &  8.2\,$\times$\,10$^{-5}$  & colour: 0.07 to 18  & n \\                                                                              
        &                         &  22$_{1,21}$\,$\rightarrow$\,21$_{1,20}$; EE &  229055.797 &   132.8       &  4.6\,$\times$\,10$^{-4}$  &             &  \\
        &                         &  22$_{2,21}$\,$\rightarrow$\,21$_{2,20}$; EE &  229055.797 &   132.8       &  4.4\,$\times$\,10$^{-4}$  &             &  \\
        &                         &  22$_{2,21}$\,$\rightarrow$\,21$_{1,20}$; EE &  229055.797 &   132.8       &  9.7\,$\times$\,10$^{-5}$  &            &  \\    \cline{3-8}
        &                         &  20$_{2,18}$\,$\rightarrow$\,19$_{3,17}$; AE &  218091.411 &   119.0       &  4.3\,$\times$\,10$^{-4}$  & contour: 0.07 to 9.0 by 1.0  &  \\
        &                         &  20$_{3,18}$\,$\rightarrow$\,19$_{2,17}$; AE &  218091.411 &   119.0       &  4.3\,$\times$\,10$^{-4}$  &             &  \\
        &                         &  20$_{2,18}$\,$\rightarrow$\,19$_{3,17}$; EA &  218091.448 &   119.0       &  4.3\,$\times$\,10$^{-4}$  &             &  \\
        &                         &  20$_{3,18}$\,$\rightarrow$\,19$_{2,17}$; EA &  218091.448 &   119.0       &  4.3\,$\times$\,10$^{-4}$  &             &  \\                                                                
\hline
\end{tabular}                                               
\tablefoot{a) Partially blended with CH$_3$OD (E, 5$_{0}$\,$\rightarrow$\,4$_{0}$) in the blue wing. However, to map the CH$_3$OCH$_3$ emission, we avoided the contribution of CH$_3$OD by selecting an appropriate interval of velocities (see Fig.\,\ref{fig_lines_1}). \\
b) \citet{Cernicharo_2016}.\\
c) Partially blended with the red wing of a CH$_2$DCN line (14$_{6,8}$\,$\rightarrow$\,13$_{6,7}$ \& 14$_{6,9}$\,$\rightarrow$\,13$_{6,8}$) in the EG and ET peaks.\\
d) \citet{Tercero_2015}.\\
e) Blended with the red wing of a U line at 228408 MHz (assuming a $v_{\rm LSR}$ of 7.5\,km\,s$^{-1}$) in the EG and ET peaks.\\
f) \citet{Tercero_2015}; partially blended with some emission from extreme velocities of $^{34}$SO$_2$ (6$_{3,3}$\,$\rightarrow$\,6$_{2,4}$) 
and SO$_2$ (26$_{3,23}$\,$\rightarrow$\,25$_{4,22}$) in the EG and ET peaks. Nevertheless, the emission at MF peak is not blended (see Fig.\,\ref{fig_lines_1}).\\
g) Blended with some emission from extreme velocities of SO$^{18}$O (12$_{1,12}$\,$\rightarrow$\,11$_{0,11}$), CH$_2$DCN (13$_{8,5}$\,$\rightarrow$\,12$_{8,4}$ \& 
13$_{8,6}$\,$\rightarrow$\,12$_{8,5}$), and HCOOH (10$_{3,7}$\,$\rightarrow$\,9$_{3,6}$) in the EG and ET peaks. Nevertheless, the emission at MF peak is not blended 
(see Fig.\,\ref{fig_lines_1}).\\
h) \citet{Cernicharo_2016}.\\
i) \citet{Brouillet_2015}.\\
j) \citet{Cernicharo_2016}.\\
k) \citet{Cernicharo_2016}.\\
l) \citet{Cernicharo_2016}; partially blended with the red wing of a CH$_2$DCN line (14$_{6,8}$\,$\rightarrow$\,13$_{6,7}$ \& 14$_{6,9}$\,$\rightarrow$\,13$_{6,8}$) in the EG and ET peaks.\\
m) Blended with some emission from extreme velocities of NH$_2$CHO (11$_{2,10}$\,$\rightarrow$\,10$_{2,9}$) in the EG and ET peaks.\\
n) \citet{Cernicharo_2016}.}                                              
\end{table*}                                                                                                                                                                                            

\clearpage
\section{Determination of column densities}
\label{appendix_b}

To model the emission of the studied species towards the MF, EG, and ET~peaks,
we extracted an averaged spectrum 
over 5\,$\times$\,5 pixels (1$''$$\times$1$''$) around the emission peaks mentioned above, and using MADEX we calculated the synthetic
spectrum between 213.7\,GHz and 246.7\,GHz for the different species and positions
according to the physical parameters collected in Table\,\ref{table_cd}.
Owing to the lack of collisional rates for most of the studied species, we
used the local thermodynamic equilibrium (LTE) approximation. Taking into account the
physical conditions of the considered components of the cloud
(see Table\,\ref{table_cd}), we expect  this approximation to work reasonably
well.

For each spectral
component and species, we assumed uniform physical conditions of the cloud:
line width, radial velocity, and a source size slightly larger than the spatial resolution of these observations 
(3$''$). 
Therefore, we left only the rotational temperature and the molecular column
density as free parameters. However,
as a first step, we relied on previous works (see e.g. \citealt{Tercero_2015,Cernicharo_2016})
to fix the rotational temperature at 150\,K.
Then we varied the column density to obtain an agreement in line intensity between model and observations
better than $\sim$30\% for all features of the studied species present in the frequency range of the ALMA SV data.
Lines for which this condition was not fulfilled
were revised searching for possible blendings with features from other
molecular species. Finally, we adopted the model parameters if all spectral features that did not match with the
synthetic spectrum were blended. We marked as the upper limit the column density results
obtained with less than five unblended lines.

For certain species, 
some mismatches between model and data could not be explained due to line blending.
We noted that lines which showed these discrepancies were high- and/or low-excitation lines. 
In these cases, we 
introduced a hotter (at 250\,K) and/or colder (at 50\,K)
component to properly fit all unblended lines by varying the
column densities of the different temperature components and following the criteria
discussed above. The derived column density shown in Table\,\ref{table_cd} is the
sum of the values obtained for the different temperature components.

In addition, for the most abundant species other components at different radial velocities have been included
to reproduce the observed line profiles (see Table\,\ref{table_cd}). We consider that
these extra components come from adjacent parts of the region that partially overlap with the emission from the
considered emission peaks.

\begin{table*}
\centering
\caption{Molecular column densities and abundances.}
\label{table_cd}
\rotatebox{90}{
\resizebox{1.3\textwidth}{!}{
\begin{tabular}{l l c c c c c c c c c c c@{\vrule height 10pt depth 5pt width 0pt}}
\hline\hline
 &  & \multicolumn{3}{c}{\bf MF Peak$^{(a)}$} \rule[0.15cm]{0cm}{0.2cm}\ & &  \multicolumn{3}{c}{\bf EG Peak$^{(b)}$} \rule[0.2cm]{0cm}{0.2cm}\ &  & \multicolumn{3}{c}{\bf ET Peak$^{(c)}$} \rule[0.15cm]{0cm}{0.2cm}\  \\ 
 
  &  & \multicolumn{3}{c}{$\upalpha_{\rm J2000}$\,=\,05$^{\rm h}$\,35$^{\rm m}$\,14.102$^{\rm s}$, $\updelta_{\rm J2000}$\,=\,$-$05$^{\circ}$\,22$'$\,36.84$''$} \rule[0.15cm]{0cm}{0.2cm}\ & &  \multicolumn{3}{c}{$\upalpha_{\rm J2000}$\,=\,05$^{\rm h}$\,35$^{\rm m}$\,14.474$^{\rm s}$, $\updelta_{\rm J2000}$\,=\,$-$05$^{\circ}$\,22$'$\,33.12$''$} \rule[0.2cm]{0cm}{0.2cm}\ &  & \multicolumn{3}{c}{$\upalpha_{\rm J2000}$\,=\,05$^{\rm h}$\,35$^{\rm m}$\,14.440$^{\rm s}$, $\updelta_{\rm J2000}$\,=\,$-$05$^{\circ}$\,22$'$\,34.74$''$} \rule[0.15cm]{0cm}{0.2cm}\  \\ 
  
&  & \multicolumn{3}{c}{$v_{\rm lsr}$\,=\,7.5\,km\,s$^{-1}$; $\Delta v$\,=\,2\,km\,s$^{-1}$; $N$(H$_2$)=8.2$\times$10$^{23}$\,cm$^{-2}$} \rule[0.15cm]{0cm}{0.2cm}\ & &  \multicolumn{3}{c}{$v_{\rm lsr}$\,=\,7.5$-$8.0\,km\,s$^{-1}$; $\Delta v$\,=\,3\,km\,s$^{-1}$; $N$(H$_2$)=2.4$\times$10$^{24}$\,cm$^{-2}$} \rule[0.2cm]{0cm}{0.2cm}\ &  & \multicolumn{3}{c}{$v_{\rm lsr}$\,=\,7.5$-$8.0\,km\,s$^{-1}$; $\Delta v$\,=\,3\,km\,s$^{-1}$; $N$(H$_2$)=1.8$\times$10$^{24}$\,cm$^{-2}$} \rule[0.15cm]{0cm}{0.2cm}\  \\ \cline{3-5} \cline{7-9}   \cline{11-13}
 
 & &  $T_{\rm rot}$  &  $N$ & $N$/$N$(H$_2$) & & $T_{\rm rot}$  &  $N$ & $N$/$N$(H$_2$) & & $T_{\rm rot}$  &  $N$ & $N$/$N$(H$_2$)  \\
& &  [K]  & $\times$ 10$^{15}$ [cm$^{-2}$] & $\times$ 10$^{-9}$ &  &  [K]  & $\times$ 10$^{15}$ [cm$^{-2}$] & $\times$ 10$^{-9}$ & & [K]  & $\times$ 10$^{15}$ [cm$^{-2}$] & $\times$ 10$^{-9}$ \\

\hline
{\bf A.}  & CH$_3$OH                &  50\,-\,150\,-\,250  &   1080         & 1305            & &  50\,-\,150\,-\,250   &  1260         & 540           &  &  50\,-\,150\,-\,250    &  3240          & 1800  \\
{\bf B.}  & HCOOCH$_3$              &  50\,-\,150\,-\,250  &   240          & 293             & &  150                  &  60           & 25            &  &  50\,-\,150\,-\,250    &  240           & 133 \\
{\bf C.}  & CH$_3$OCH$_3$           &  50\,-\,150\,-\,250  &   100          & 122             & &  150\,-\,250          &  60           & 25            &  &  50\,-\,150\,-\,250    &  110           & 61  \\
{\bf D.}  & CH$_2$OCH$_2$           &  150                 &   2.7          & 3.3             & &  150                  &  2.0          & 0.8           &  &  150                   &  3.3           & 1.8 \\
{\bf E.}  & CH$_3$COOCH$_3$         &  150                 &   6.0          & 7.3             & &  150                  &  3.1          & 1.3           &  &  150                   &  6.0           & 3.3 \\
{\bf F.}  & HCOOCH$_2$CH$_3$        &  150                 &   3.0          & 3.7             & &  150                  &  2.0          & 0.8           &  &  150                   &  4.0           & 2.2  \\
{\bf G.}  & CH$_3$CH$_2$OCH$_3$     &  150                 &   3.0          & 3.7             & &  150                  &  $\leq$\,2.0  &  $\leq$\,0.8  &  &  150                   & $\leq$\,2.0    & $\leq$\,1.1 \\
{\bf H.}  & HCOOH                   &  150                 &  0.3           & 0.4             & &  150                  &   1.5         & 0.6           &  &  150                   &  1             & 0.6 \\
{\bf I.}  & OHCH$_2$CH$_2$OH        &  150                 &  $\leq$\,0.3   & $\leq$\,0.4     & &  150                  &   5.3         & 2.2           &  &  150                   &  3.2           & 1.8 \\
{\bf J.}  & CH$_3$COOH              &  150                 &  $\leq$\,0.4   & $\leq$\,0.5     & &  150                  &   4.5         & 1.9           &  &  150                   &  2             & 1.1  \\
{\bf K.}  & CH$_3$CH$_2$OH          &   50\,-\,150\,-\,250 &  4.0           & 4.9             & &  50\,-\,150\,-\,250   &   7.0         & 2.9           &  &  50\,-\,150\,-\,250    &  64            & 36 \\
{\bf L.}  & CH$_3$OCH$_2$OH         &  150                 &  15            & 18              & &  150                  &   120         & 50            &  &  150                   &  200           & 111 \\
{\bf M.}  & OHCH$_2$CHO             &  150                 &  $\leq$\,0.1   & $\leq$\,0.1     & &  150                  &  $\leq$\,0.3  & $\leq$\,0.1   &  &  150                   & $\leq$\,0.3    & $\leq$\,0.2 \\
{\bf N.}  & CH$_3$COCH$_3$          &  150                 &  0.5           & 0.6             & &  50\,-\,150\,-\,250   &   4.0         & 1.7           &  &  50\,-\,150\,-\,250    &  5.0           & 2.8 \\

\hline

\end{tabular}
}}
\rotatebox{90}{
\tablefoot{\tiny Rotational temperatures ($T_{\rm rot}$), column densities ($N$), and abundances ($N$/$N$(H$_2$)) inferred towards the MF, EG, and ET peaks in Orion~KL.\\
(a): Position in agreement with the MF1 peak of \citet{Favre_2011a} and position A of \citet{Tercero_2015}.\\
(b): Position in agreement with the EG peak of \citet{Brouillet_2015} and \citet{Favre_2017}.\\
(c): Position in agreement with the MF2 peak of \citet{Favre_2011a} and position B of \citet{Tercero_2015}.\\
\textbf{A}: model for the $a$-type lines ($b$-type lines are optically thick); A\,+\,E states; $^{12}$C/$^{13}$C\,=\,45 \citep{Tercero_2010}; another component has been included to properly fit the observed line profiles: \textbf{MF~peak}: $v_{\rm LSR}$\,=\,9.5\,km\,s$^{-1}$, $\Delta v$\,=\,3\,km\,s$^{-1}$, $T_{\rm rot}$\,=\,150\,K,
$N$\,=\,1.4\,$\times$\,10$^{18}$\,cm$^{-2}$; \textbf{EG~peak}: $v_{\rm LSR}$\,=\,6.0\,km\,s$^{-1}$, $\Delta v$\,=\,8\,km\,s$^{-1}$, $T_{\rm rot}$\,=\,150\,K,
$N$\,=\,1.8\,$\times$\,10$^{18}$\,cm$^{-2}$; \textbf{ET~peak}: $v_{\rm LSR}$\,=\,6.0\,km\,s$^{-1}$, $\Delta v$\,=\,8\,km\,s$^{-1}$, $T_{\rm rot}$\,=\,150\,K,
$N$\,=\,9.0\,$\times$\,10$^{17}$\,cm$^{-2}$.\\
\textbf{B}: model for the $b$-type lines ($a$-type lines are optically thick); A+E states; another component has been included to properly fit the observed line profiles: \textbf{MF~peak}: $v_{\rm LSR}$\,=\,9.0\,km\,s$^{-1}$, $\Delta v$\,=\,4\,km\,s$^{-1}$, $T_{\rm rot}$\,=\,150\,K, $N$\,=\,1.0\,$\times$\,10$^{17}$\,cm$^{-2}$;
\textbf{EG~peak}: $v_{\rm LSR}$\,=\,6.0\,km\,s$^{-1}$, $\Delta v$\,=\,8\,km\,s$^{-1}$, $T_{\rm rot}$\,=\,150\,K, $N$\,=\,6.0\,$\times$\,10$^{16}$\,cm$^{-2}$.\\
\textbf{C}: AA\,+\,AE\,+\,EA\,+\,EE states; another component has been included to properly fit the observed line profiles: \textbf{MF~peak}: $v_{\rm LSR}$\,=\,9.0\,km\,s$^{-1}$,
$\Delta v$\,=\,4\,km\,s$^{-1}$, $T_{\rm rot}$\,=\,150\,K, $N$\,=\,3.0\,$\times$\,10$^{16}$\,cm$^{-2}$.\\
\textbf{D}: ortho\,+\,para symmetry.\\
\textbf{E}: AA\,+\,AE\,+\,EA\,+\,E3\,+\,E4 states.\\
\textbf{F}: trans\,+\,gauche conformers.\\
\textbf{G}: AA\,+\,AE\,+\,EA\,+\,EE\,+\,EE' states.\\
\textbf{H}: trans conformer.\\
\textbf{I}: (anti-gauche)\,+\,(gauche-gauche) conformers.\\
\textbf{J}: A\,+\,E states.\\
\textbf{K}: trans\,+\,gauche conformers; another component has been included to properly fit the observed line profiles: \textbf{MF~peak}: $v_{\rm LSR}$\,=\,9.0\,km\,s$^{-1}$,
$\Delta v$\,=\,4\,km\,s$^{-1}$, $T_{\rm rot}$\,=\,150\,K, $N$\,=\,1.5\,$\times$\,10$^{16}$\,cm$^{-2}$.\\
\textbf{N}: AA\,+\,AE\,+\,EA\,+\,EE states; another component has been included to properly fit the observed line profiles: \textbf{EG~peak}: $v_{\rm LSR}$\,=\,6.0\,km\,s$^{-1}$,
$\Delta v$\,=\,8\,km\,s$^{-1}$, $T_{\rm rot}$\,=\,150\,K, $N$\,=\,3.0\,$\times$\,10$^{15}$\,cm$^{-2}$; \textbf{ET~peak}: $v_{\rm LSR}$\,=\,6.0\,km\,s$^{-1}$,
$\Delta v$\,=\,8\,km\,s$^{-1}$, $T_{\rm rot}$\,=\,150\,K, $N$\,=\,2.0\,$\times$\,10$^{15}$\,cm$^{-2}$.}
}
\end{table*}

\clearpage

\section{Complementary Figures}

\begin{figure*}[!ht]
 \vspace*{0.6cm}
\includegraphics[scale=0.63,angle=0]{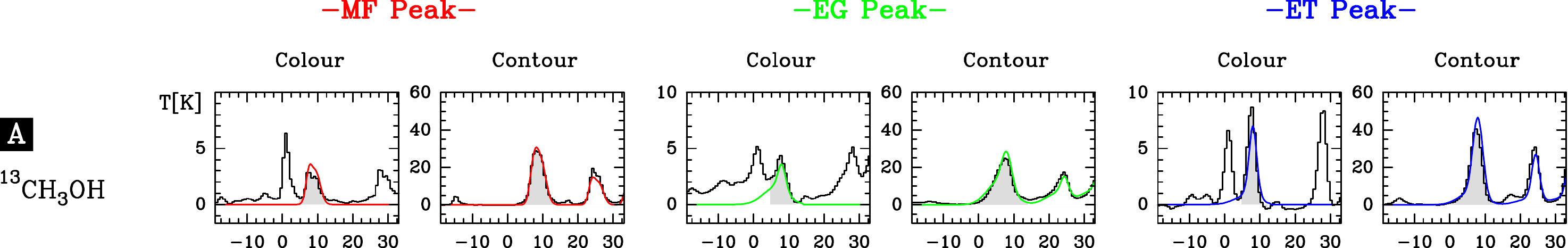}\vspace{0.5cm}\\ \vspace{0.5cm}
\includegraphics[scale=0.63,angle=0]{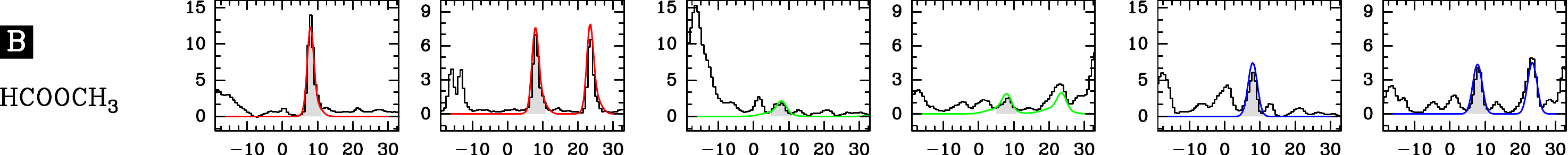}\\ \vspace{0.5cm}
\includegraphics[scale=0.63,angle=0]{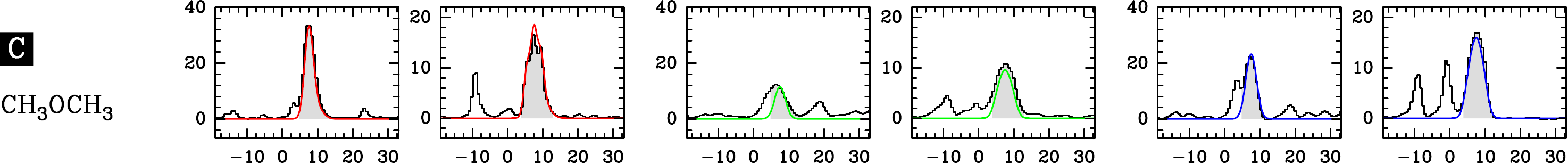}\\ \vspace{0.5cm}
\includegraphics[scale=0.63,angle=0]{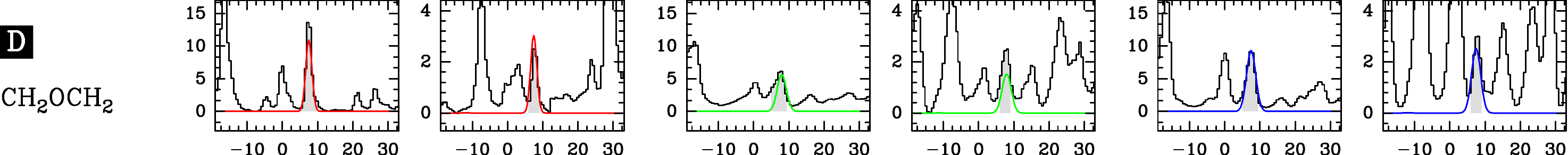}\\ \vspace{0.5cm}
\includegraphics[scale=0.63,angle=0]{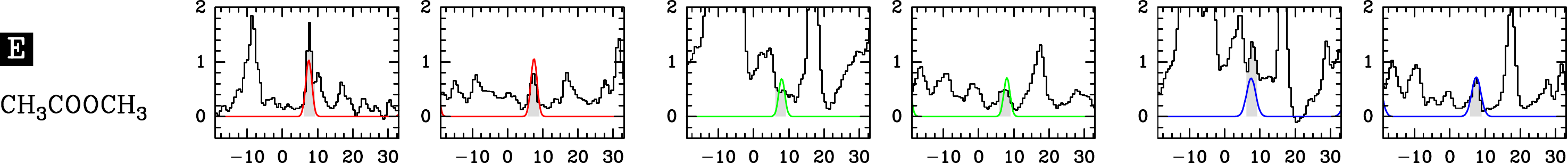}\\ \vspace{0.5cm}
\includegraphics[scale=0.63,angle=0]{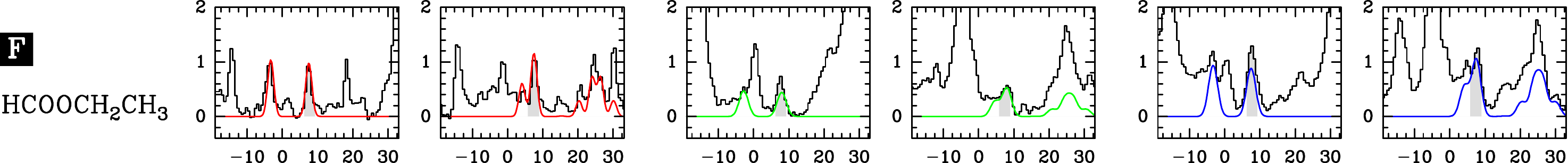}\\ 
\includegraphics[scale=0.63,angle=0]{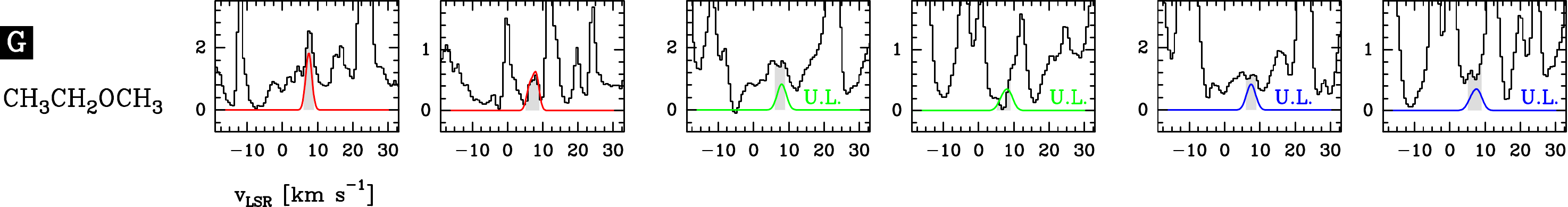}\\ 
\caption{Lines mapped in Fig.~\ref{fig_maps} (colour and contour) in the three selected positions (MF, EG, and ET peaks). The data (black histogram spectra) are from ALMA Science Verification observations. The quantum numbers of the selected lines and their spectroscopic parameters are listed in Table~\ref{table_lines}. We also show the adopted integrated area for imaging the spatial distribution of each line (Fig.~\ref{fig_maps}).The red, green, and blue curves show our best LTE models for the emission of each molecule in each position (see Table~\ref{table_cd}). The label U.L. indicates that the column density of the model is only an upper limit. The intensity scale 
has been maintained for the same line (colour or contour) in the three Orion components.}
\label{fig_lines_1}
 \end{figure*}

\clearpage

\begin{figure*}[!ht]
\setcounter{figure}{0}
 \vspace*{0.6cm}
\includegraphics[scale=0.63,angle=0]{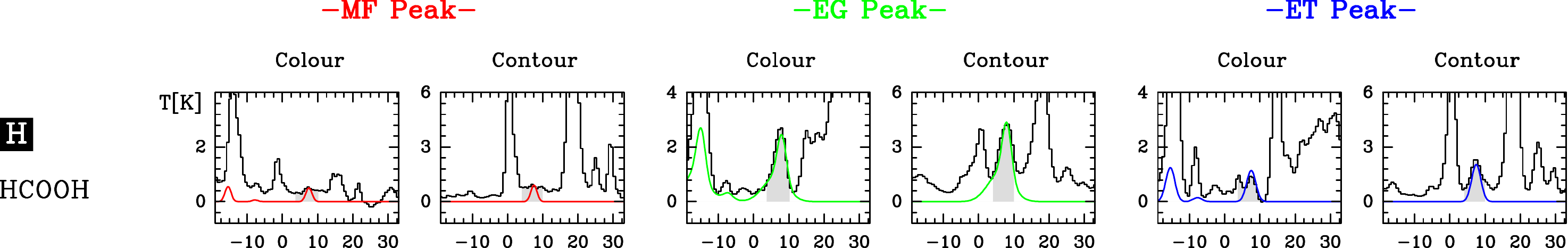}\vspace{0.5cm}\\ \vspace{0.5cm}
\includegraphics[scale=0.63,angle=0]{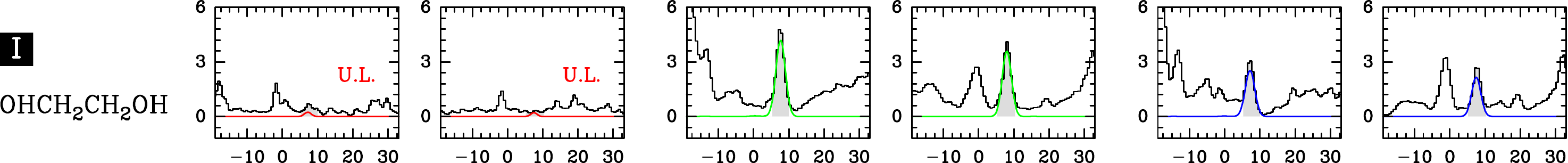}\\ \vspace{0.5cm}
\includegraphics[scale=0.63,angle=0]{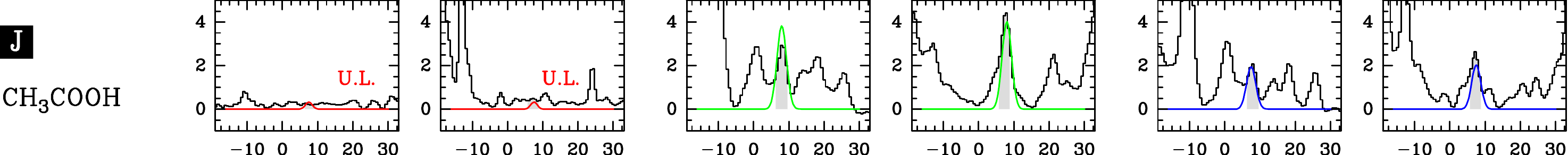}\\ \vspace{0.5cm}
\includegraphics[scale=0.63,angle=0]{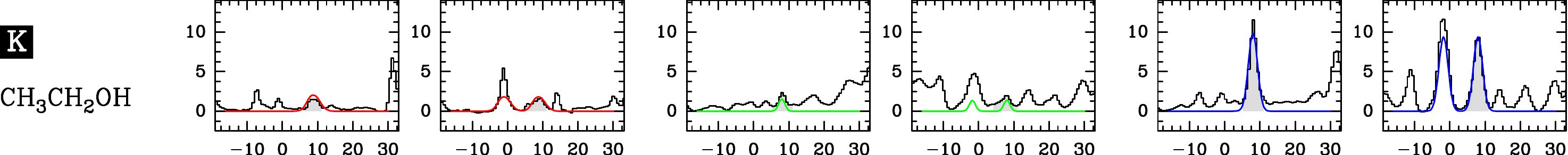}\\ \vspace{0.5cm}
\includegraphics[scale=0.63,angle=0]{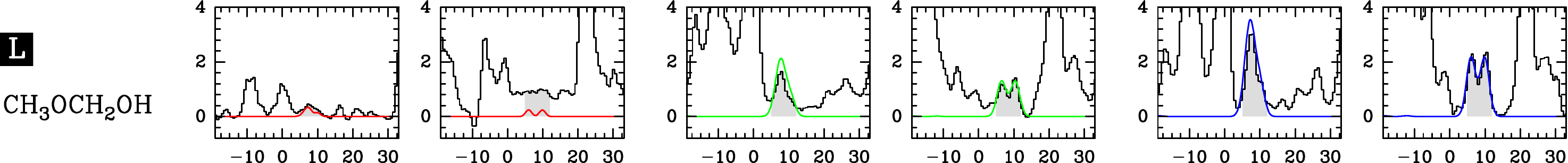}\\ \vspace{0.5cm}
\includegraphics[scale=0.63,angle=0]{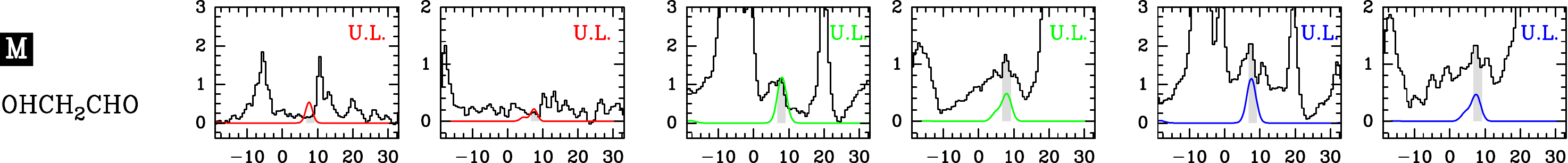}\\ \vspace{0.5cm}
\includegraphics[scale=0.63,angle=0]{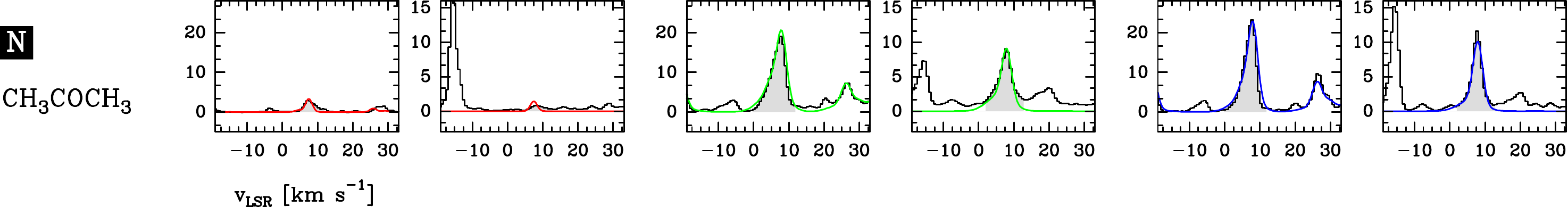}\\
\caption[]{continued.}
 \end{figure*}

\clearpage

\begin{figure*}[!ht]
\centering
 \vspace*{0.6cm}
\includegraphics[scale=0.8,angle=0]{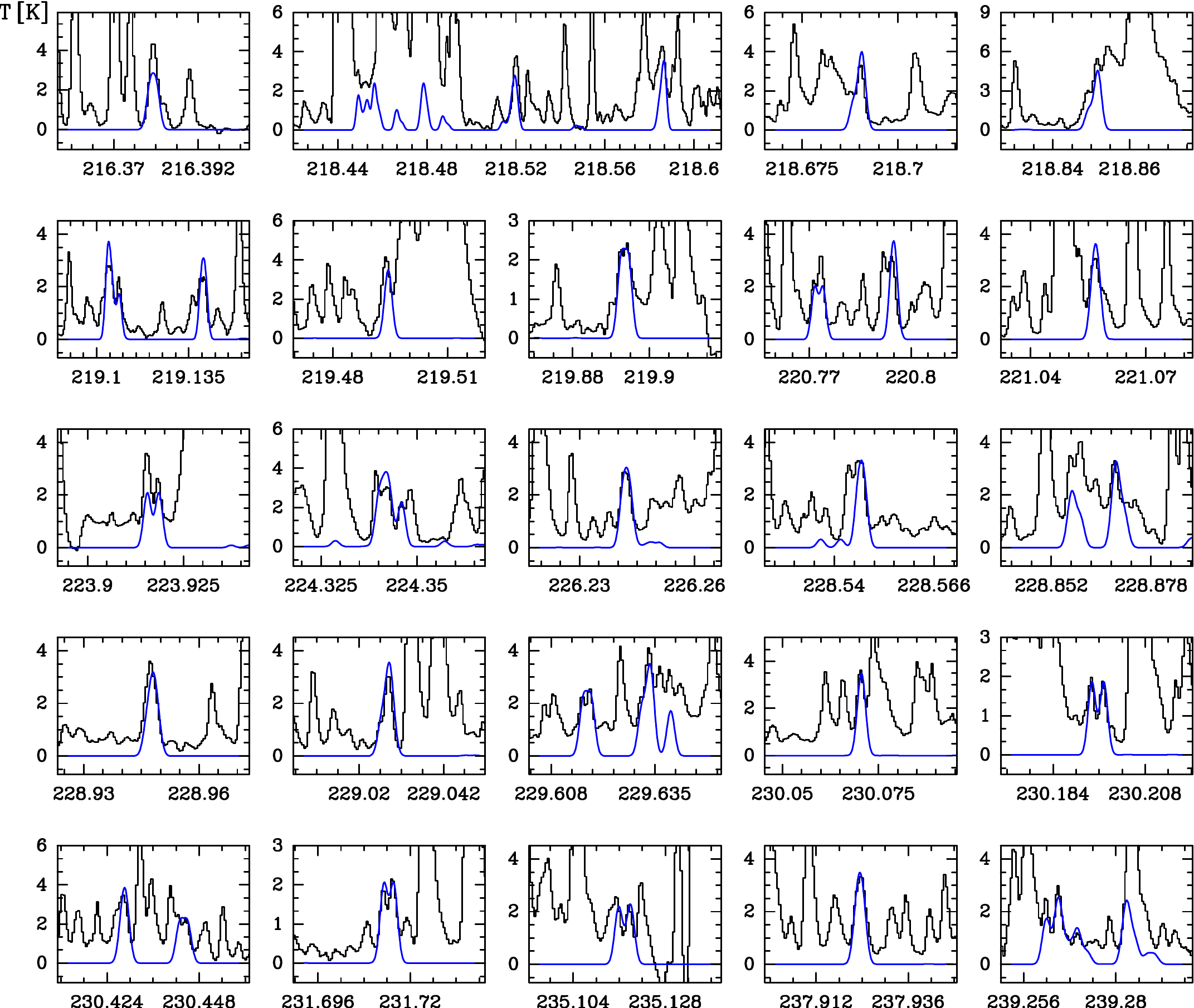}\vspace{0.45cm}
\includegraphics[scale=0.8,angle=0]{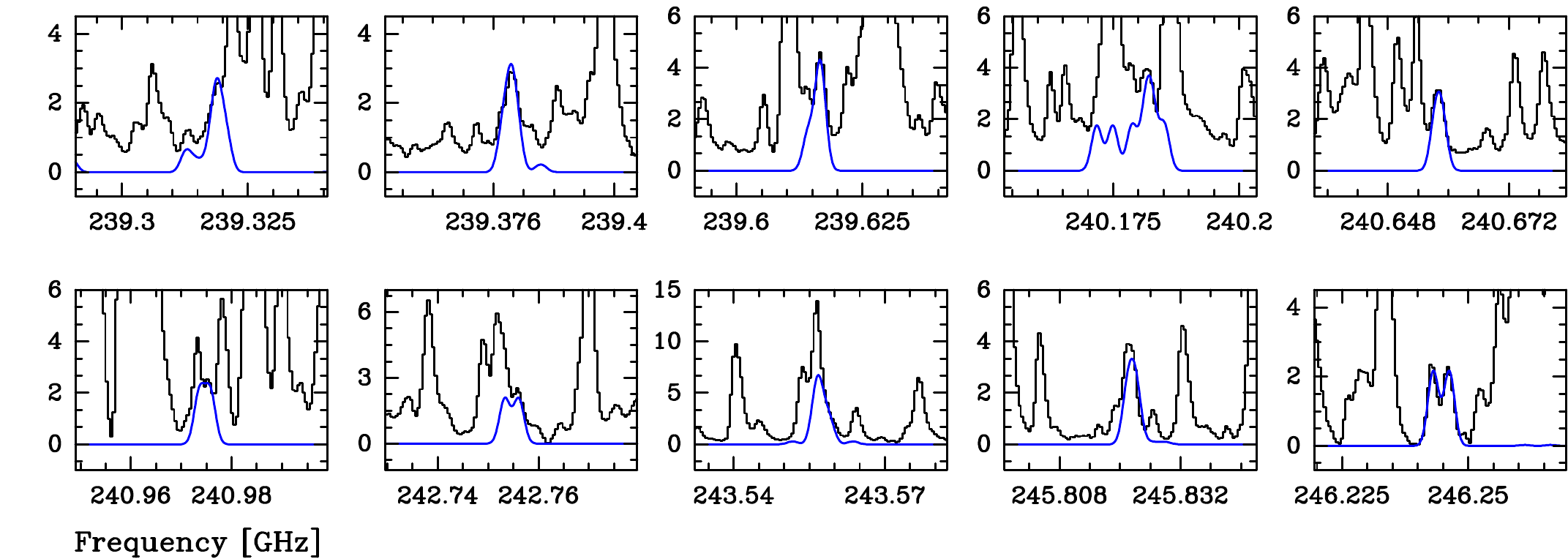}
\caption{Selected lines of CH$_3$OCH$_2$OH towards the ET peak detected with the ALMA interferometer. 
A $v_{\rm LSR}$ of +7.5\,km\,s$^{-1}$ is assumed.}
\label{fig_ch3och2oh}
\end{figure*}

\end{appendix}


\end{document}